\documentclass[conference]{IEEEtran}
\usepackage{graphicx}
\usepackage{amsmath,multicol,amsthm}
\usepackage{amsfonts}
\usepackage[dvipsnames]{xcolor}
\usepackage{url}

\usepackage[ruled]{algorithm}
\usepackage{algpseudocode}

\usepackage{ioa_code}

\algrenewcommand\algorithmiccomment[1]{\hspace{2em}{\color{blue} // \textit{#1}}}

\newcommand{\Nat}{\ifmmode \mathbb{N} \else $\mathbb{N}$ \fi}

\newcommand{\Real}{\ifmmode \mathbb{R} \else $\mathbb{R}$ \fi}

\newcommand{\Fin}[1]{\ifmmode \mathbb{F}_#1 \else $\mathbb{F}_#1$ \fi}

\newcommand{\tup}[1]{%
    \relax\ifmmode
      \langle #1 \rangle%
    \else
        $\langle$#1$\rangle$%
    \fi
}

\newcommand{\act}[1]{%
    \relax\ifmmode
        \mathord{\mathcode`\-="702D\sf #1\mathcode`\-="2200}%
    \else
        $\mathord{\mathcode`\-="702D\sf #1\mathcode`\-="2200}$%
    \fi
}

\newcommand{\remove}[1]{}

\newcommand{\mtype}[1]{\text{{\sc #1}}}

\newcommand{\WRP}{\par\qquad\(\hookrightarrow\)\enspace}

\newcommand{\T}{\hspace*{1em}}

\makeatletter
\def\mainlistofsymbols{
  \normalsize
  \vspace*{1.5 em}
  \@starttoc{los}
}

\def\partonelistofsymbols{
  \normalsize
  \vspace*{1.5 em}
  \@starttoc{p1los}
}

\def\parttwolistofsymbols{
  \normalsize
  \vspace*{1.5 em}
  \@starttoc{p2los}
}

\def\l@symbol#1#2{\addpenalty{-\@highpenalty} \vskip 4pt plus 2pt
{\@dottedtocline{0}{0em}{8em}{#1}{#2}}}
\makeatother

\newcommand{\newhiddensym}[2]{%
}

\newcommand{\algIOA}[2]{\ifmmode{\text{#1}_{#2}}\else{$\text{#1}_{#2}$}\fi}
\newcommand{\EX}{\ifmmode{\xi}\else{$\xi$}\fi}
\newcommand{\EXF}{\ifmmode{\phi}\else{$\phi$}\fi}
\newcommand{\inter}[1]{
	\ifmmode{\left(\bigcap_{\mathcal{Q}\in#1}\mathcal{Q}\right)}
	\else{$\left(\bigcap_{\mathcal{Q}\in#1}\mathcal{Q}\right)$}
	\fi
}
\newcommand{\idSet}{\mathcal{I}}
\mathchardef\mhyphen="2D
\newcommand{\pr}{p}
\newcommand{\vid}[1]{\ifmmode{\nu_{#1}}\else{$\nu_{#1}$}\fi}
\newcommand{\seen}{\ifmmode{seen}\else{$seen$}\fi}
\newcommand{\valSet}{{\mathcal V}}
\newcommand{\msgSet}{M}
\newcommand{\maxts}[1]{\ifmmode{maxTS_{#1}}\else{$maxTS_{#1}$}\fi}
\newcommand{\maxtag}[1]{\ifmmode{maxTag_{#1}}\else{$maxTag_{#1}$}\fi}
\newcommand{\maxpair}[1]{\ifmmode{maxMPair_{#1}}\else{$maxMPair_{#1}$}\fi}
\newcommand{\mintag}[1]{\ifmmode{minTag_{#1}}\else{$minTag_{#1}$}\fi}
\newcommand{\maxps}{\ifmmode{maxPS}\else{$maxPS$}\fi}
\newcommand{\conftg}[1]{\ifmmode{confirmed_{#1}}\else{$confirmed_{#1}$}\fi}
\newcommand{\maxconftag}{\ifmmode{\ms{maxCT}}\else{$maxCT$}\fi}

\newcommand{\hash}[1]{\mathcal{H}^{#1}}

\newcommand{\vertexSet}{V}

\renewcommand{\msgSet}{\mathcal{M}}

\newcommand{\optpp}{{\sc OptimumP2P}}

\newcommand{\gossipsub}{Gossipsub}
\newcommand{\floodsub}{Floodsub}

\algblockdefx[Operation]{Operation}{EndOperation}%
[2]{{\bf operation} $\act{#1}$(#2)}%
{{\bf end operation}}
\algblockdefx[Procedure]{Procedure}{EndProcedure}%
[2]{{\bf procedure} $\act{#1}$(#2)}%
{{\bf end procedure}}
\algblockdefx[Receive]{Receive}{EndReceive}%
[2]{{\bf Upon receive} (#1)$_{\text{ #2 }}${\bf from} $q$}%
{{\bf end receive}}

\newcommand{\nn}[1]{{#1}}

\newcommand{\myparagraph}[1]{\smallskip\noindent{\textbf{#1}}}
\renewcommand{\pr}{v}

\title{\optpp{}: Fast and Reliable Gossiping in P2P Networks}

\author{
\IEEEauthorblockN{
Nicolas Nicolaou\textsuperscript{1},
Onyeka Obi\textsuperscript{1},
Aayush Rajasekaran\textsuperscript{1},
Alejandro Bergasov\textsuperscript{1},
Aleksandr Bezobchuk\textsuperscript{1}, \\
Kishori M. Konwar\textsuperscript{1},
Michael Meier\textsuperscript{1},
Santiago Paiva\textsuperscript{1},
Har Preet Singh\textsuperscript{1},
Swarnabha Sinha\textsuperscript{1}, \\
Sriram Vishwanath\textsuperscript{2}, and 
Muriel M\'edard\textsuperscript{1}
}
\IEEEauthorblockA{\textsuperscript{1}\texttt{\{nicolas,oni,aayush,alejandro,bez,kkonwar,mike, }}
\IEEEauthorblockA{\textsuperscript{1}\texttt{santiago,harpreet,swarna,mmedard\}@getoptimum.xyz}}
\IEEEauthorblockA{\textsuperscript{1}Optimum, Cambridge, MA, USA}
\IEEEauthorblockA{\textsuperscript{2}\texttt{sriram@ece.gatech.edu}}
\IEEEauthorblockA{\textsuperscript{2}Georgia Tech, GA, USA}
}

\begin{document}
%\date{}

\maketitle

\begin{abstract}
    Gossip algorithms are pivotal in the dissemination of information within decentralized systems. Consequently, numerous gossip libraries have been developed and widely utilized especially in blockchain protocols for the propagation of blocks and transactions.
    A well-established library is \emph{libp2p}, which provides two gossip algorithms: {\em floodsub} and {\em gossipsub}. These algorithms enable the delivery of published messages to a set of peers. In this work 
    %in progress (WIP), 
    we aim to enhance the performance and reliability of \emph{libp2p} by introducing \optpp{}, a novel gossip algorithm that leverages the capabilities of Random Linear Network Coding (RLNC) to expedite the dissemination of information in a peer-to-peer (P2P) network while ensuring reliable delivery, even in the presence of malicious actors capable of corrupting the transmitted data. 
    Preliminary research from the Ethereum Foundation has demonstrated the use of RLNC in the significant improvement in the block propagation  time~\cite{ethresearch-rlnc-artice}. Here we present extensive evaluation results 
    both in simulation and real-world environments that demonstrate the performance gains of \optpp{} over the 
    \gossipsub{} protocol.
    % RLNC has been shown to be order optimal for an arbitrary graph in terms of completion of propagation for both pull and push models. We present here \optpp{} a gossip protocol based on RLNC. 
\end{abstract}

%\begin{IEEEkeywords}
%	atomic storage; erasure codes; fault tolerance;
%\end{IEEEkeywords}
%

\section{Introduction}
\label{sec:intro}
%\kk{We need to thank Google Cloud providers for the grant}
Gossip algorithms, also known as epidemic protocols~\cite{birman2007promise, eugster2004epidemic,kermarrec2007gossiping}, are a class of decentralized communication strategies used in distributed systems to disseminate information efficiently and robustly across a network. In these algorithms, nodes periodically exchange information with a randomly selected subset of neighboring nodes, mimicking the spread of gossip in social networks. In a sense, this is also strongly related to epidemics, by which a disease is spread by infecting members of a group, which in turn can infect others. This probabilistic approach ensures that information propagates rapidly and reliably, even in large-scale or dynamically changing networks, without requiring centralized coordination or global knowledge of the system topology. These characteristics make gossip algorithms inherently fault-tolerant, scalable, and adaptable, making them ideal for applications such as distributed databases, consensus protocols, and peer-to-peer networks. 

Consequently, gossip protocols attracted the attention and were 
widely adopted in implementations of blockchain
systems. They serve an efficient and reliable solution for various tasks, including transaction 
and block propagation as seen in Bitcoin~\cite{N08bitcoin} and Ethereum~\cite{buterin2014ethereum}, peer discovery (e.g.,Ethereum~\cite{buterin2014ethereum}), reaching consensus (e.g., Tendermint~\cite{buchman2016tendermint}), and state synchronization (e.g., Hyperledger Fabric~\cite{androulaki2018hyperledger}) among others. 

However, blockchain implementations often operate on top of a permissionless, asynchronous, message-passing network, susceptible to unpredictable delays and node
failures. Therefore, improper use of gossiping approaches may lead to high network
overhead and congestion, high propagation latencies, and erroneous information
propagation due to message alteration by malicious actors. \emph{libp2p}~\cite{libp2p}, is one of the latest network communication frameworks and gossip algorithms used in modern 
blockchain solutions like Ethereum 2.0~\cite{ethereum2}. \emph{libp2p} adopts two different push gossiping algorithms: {\bf \emph{\floodsub{}}} and {\bf \emph{\gossipsub{}}}. 

\floodsub{}, uses a flooding strategy where every node forwards messages to all of its neighbors. Although very efficient in discovering
the shortest path 
%to all the nodes in the network 
and very robust in delivering a
message to all the peers in the network, \floodsub{} suffered from bandwidth saturation and 
unbounded degree flooding. 

\gossipsub{} is the successor of \floodsub{}, 
which addressed the shortcomings of the initial algorithm by organizing peers into topic-based mesh, network overlay, with a target mesh degree D and utilizing control messages for reducing message duplication.
% which addressed
% the shortcomings of the initial algorithm by employing a bounded degree $D$ mesh construction 
% for the network topology, and utilizing controlled messages to
% reduce the message duplication within the network. 
Briefly, the \gossipsub{} protocol works as follows. A publisher selects $D$ peers among its peers and broadcasts its message to them. Each peer receiving a message performs preliminary validation and rebroadcasts the message to another $D$ peers. Peers exchange control messages such as {\sc iwant}, {\sc ihave} or 
{\sc idontwant} to inform their peers about their status regarding the 
propagation of a particular message. These enhancements enabled \gossipsub{}
to reduce bandwidth usage, but the introduction of the bounded degree $D$
increased the number of hops a message required to reach distant peers, resulting
in higher delivery latencies. Furthermore, similar to the \floodsub{} protocol, each 
peer forwards the full message to its peers even when other peers may already have 
received the full message, also suffering from (reduced compared to \floodsub{}) message duplication. 

\emph{So can we introduce a gossip protocol that is light in network usage and yet fast in information diffusion?} 

% \paragraph{\bf{Galois Gossiping on an asynchronous network.}}We consider a network of nodes interconnected by asynchronous communication links where each node independently and frequently generates messages. These nodes employ a gossip protocol to spread these messages throughout the network. In this scenario, gossiping entails each node transmitting its own messages, along with messages received from others, to neighboring nodes in the network. This recursive and ongoing method guarantees that, eventually, all nodes are apprised of all messages despite the asynchronous nature of their functions and the continuous creation of new data. A reliable transport protocol is crucial here, ensuring that messages are consistently delivered without loss and thus increasing fault tolerance and efficiency. 
\nn{An idea to leverage coding in network gossip was proposed in \cite{YGA22}. Luby Transform (LT) \cite{Luby2002} 
codes were used that stream algebraically coded elements from a single 
source to multiple destinations. Although efficient in direct multicast and shallow network 
topologies, the performance of LT degrades when used in highly decentralized and multi-hop topologies where the 
central code generator may need to retransmit critical coded elements that may be lost 
during multi-hop relays.}
% there, and, given that LT codes require original data as part of the encoding process, this choice of coding does not allow for recoding without decoding.

%\paragraph{Contributions.}
In this work we propose the use of \textbf{Random Linear Network Coding (RLNC)}~\cite{RLNC2006} 
for message broadcast. 
RLNC is a technique used in communication networks to enhance data transmission efficiency and robustness. In RLNC, data packets are encoded as random linear combinations of original packets over a finite field, typically \( \mathbb{F}_{2^m} \). This approach allows intermediate nodes in the network to mix packets without decoding (aka \textit{recode}), and the receiver can recover the original data by solving a system of linear equations once enough linearly independent combinations are received. 
 RLNC leverages the algebraic properties of finite fields, that are deeply rooted  in \textit{Galois theory}~\cite{lang2002algebra}, to ensure that the encoding and decoding processes are both efficient and probabilistically reliable. Using RLNC for gossip was introduced in~\cite{DMC05, DMC06}, which showed that optimum $O(n)$ dissemination of $k$ messages is possible for both pull and push. The analysis was refined by \cite{BAL10}, which considered pull, push and exchange, using Jackson networks with network coding \cite{Jac63} in a manner akin to \cite{LUN20083}.  Other settings, such as nodes with mobility \cite{HV14, WL18, Zhangetal13}, broadcast edges (equivalent to hyperedges) \cite{FR13, GGP19}  or correlated data \cite{CHAM15}, have been considered. Some initial results with large transfer of files were reported in \cite{LN11}.
%TCF09}. 
Probably the most significant results are those that have shown, by using projection analysis \cite{Hae11} to consider the stopping time of gossip with RLNC, that, beyond order optimality in $n$, RLNC gossip achieves  “perfect pipelining” \cite{Hae16}. The stopping time converges with high probability in optimal time, namely in time of $O(k + T)$, where $k$ is the number of messages and $T$  the dissemination time of a single message. Note that the general problem of network coding dissemination is hard to analyze when we do not use a large field size \cite{GLS22}.
 % \nn{A similar idea to leverage coding in network gossip was proposed in \cite{YGA22}, although Luby Transform (LT) codes were chosen there, and, given that LT codes require original data as part of the encoding process, this choice of coding does not allow for recoding without decoding.} 

While the above results point to the potential benefit for using RLNC in gossip, in order for RLNC to be deployed in current decentralized systems, it requires the design of a full protocol.  \optpp{} is  a novel gossip mechanism based on RLNC, hence the term \textit{Galois Gossip}, which significantly enhances the spread of information across the network.
%, resembling the swift and effective dissemination typical of social exchanges. 
\nn{More precisely, as any network coding algorithm, RLNC allows the publisher
to split a message into coded fragments (\textit{shards}) and send a subset 
%single shard 
(or a linear combination) of shards 
-- instead of the 
full message -- to each of its peers. In turn, peers can forward linear combinations of shards they receive to their own peers.} This approach has dual benefit: 
\begin{enumerate}
    \item  {\bf Faster Network Coverage:} it allows each peer to reach more peers for the same amount of data sent in full message counterparts (e.g., \gossipsub{}), and
    \item {\bf Message Duplication Reduction:} as peers receive different shards in parallel from different peers which combine to decode the original message.
\end{enumerate}

Essentially, \optpp{} allows peers to spread information \textit{piece by piece}, as oppose to traditional gossip approaches that 
broadcast full information between any pair of peers. 
% In addition, RLNC provides 
% the ability to detect when any of the shreds received is maliciously manipulated. 
% We use this knowledge to develop an algorithm that detects and avoids malicious 
% actors in the network, enhancing the fault-tolerance of the system to Byzantine failures. 
Overall \optpp{} is a new protocol implemented within \textit{libp2p} 
%—a peer-to-peer network transport framework—
aiming to decrease latency, enhance fault tolerance, and optimize bandwidth usage.
In the rest of the document we present the \optpp{} protocol and extensive 
experiments we conducted to compare the protocol's performance with \gossipsub{}, 
both in a simulation and real-world environments.

\section{System Model}
\label{sec:model}
\optpp{} aims to built a gossip service on top of a set of asynchronous, message-passing,
network processes, we refer to as \textit{peers}, a subset of which may fail arbitrarily. 
Each peer has a unique identifier from a set $\idSet$ and has access to a local clock which is not synchronized across peers. 
% Clocks between peers are not synchronized and peers may take execution steps 
% in an asynchronous fushion, i.e., with arbitrary execution delays. 

\myparagraph{\bf{Gossip Service}:} We assume a gossip service where peers may perform two primitive operations: (i) $\act{publish}(m)_p$ operation where a peer $p\in\idSet$ requests the dissemination of a message $m$ among the peers in $\idSet$, and (ii) a $\act{deliver}(m)_p$ operation that delivers a message $m$ to a peer $p\in\idSet$.
  From a user point of view a Gossip Service is defined by the following properties. 
 \begin{itemize}
 \item \textit{Validity}: if a peer publishes a message $m$, then 
 $m$ is eventually delivered at every correct peer.
\item
  \textit{Integrity}: a message $m$ is delivered by a peer, if and only if $m$ was previously published by some peer.
\end{itemize}

% An application built on a Gossip service interfaces with the Gossip component in two
% pieces of functionality, \emph{Publishing} and \emph{Delivering}. The application can submit
% messages to be Published by the Gossip service, and waits for the Gossip
% service to Deliver it messages that have been published by other peers.

% A Gossip service has the following semantics:
% \begin{itemize}
%     \item If Node A Publishes a message M, its contents will be eventually Delivered by every correct Node B
%     \item If Node A Delivers a message M, there must exist a Node B that Published M
% \end{itemize}
% \paragraph{\bf{System Model.}}
% The network is asynchronous and reliable. Each node has access to its local clock and all progress at the same rate, although they may not be synchronized. Local computations ignore and assumed to take negligible time.
% We assume the communication between any two pair of node to be \emph{asynchronous}, which  does not require the participating systems to operate simultaneously. This method is more flexible, allowing data transmission at any time, regardless of the receiver's state. Each method has its advantages and presents different engineering problems.

\myparagraph{\bf{Communication Graph}:}
Peers communicate through \textit{asynchronous} channels. We assume two primary types of channels: reliable and unreliable.
We represent our communication network by a directed graph
$G = (V, E)$, where $V\subseteq\idSet$ is the set of vertices, representing the set of peers that participate in the service, and $E$ the set of edges, or a set of links such that information can be reliably communicated from peer $u$ to $v$ for each $(u, v) \in E$.  Each link $e$, $e \in E$ is associated with a non-negative number $w_e$ representing the transmission capacity of the link in bit per unit time. 
The nodes $u$ and $v$ are referred to as \emph{origin} and \emph{destination}, respectively, of the link $(u, v) \in E$. 
%The information transmitted on a link $(u, v)$ is obtained as a coding function of information previously received at $v$.

\myparagraph{\bf{Messages}:} Each published message gets a unique identifier from a set $\msgSet$,
and contains a stream of bytes we refer to as the content of the message with a value $\mathbf{v} \in \valSet$.\footnote{Note that the contents of a message can also be made unique by adding a random number from a large prime field, e.g., $\mathbb{F}_{2^q}$} 
\nn{ We consider the use of \emph{collision-resistant} cryptographic hash function which we denote by $\hash{}: \{0, 1\}^* \rightarrow \{0, 1\}^n$ \cite{Karger1997} 
for the generation of message identifiers in $\msgSet$.}
A $\act{publish}$ operation aims to propagate the 
contents of a message $m\in\msgSet$, while a $\act{deliver}$ operation aims to 
retrieve the contents of $m$ and return them to the receiving peer. 

\myparagraph{\bf Encoding/Decoding with RLNC:}
% Each message $\mathbf{v} \in \valSet$ is identified with a unique $id$ irrespective of the content and identical message contents are made unique by adding a random number from a large prime field. Moreover, in the discussion below we will refer to a message by its $id$.
%The Galois gossip protocol uses 
\nn{We use Random Linear Network Codes (RLNC) over a finite field $\mathbb{F}_{2^q}$, to encode the contents of a message $m\in\msgSet$. In particular, for a given parameter $k$, a peer encodes the contents $\mathbf{v} \in\valSet$, using RLNC, to $k*p$ coded elements (or \textit{shards}), for some $p\geq 1$. Subsequently, the encoder or any other peer in the network may linearly combine any 
subset of shards to derive new linear combinations (i.e., new shards). 
Any $k$ of the generated shards is sufficient to decode the value $\mathbf{v}$.  We assume that $k$ does not vary from message to message.  
%In other words, we choose $k$ large enough that it will be sufficient for any number of different messages that we wish to send. 
For encoding, we do the following: $(i)$ divide $\mathbf{v}$ into a vector of $k$ elements $(v_1, v_2, \ldots, v_k)$; $(ii)$ select a matrix $\mathbf{A}$ of coefficients at random from the finite field $\mathbb{F}_{2^q}$ such that $\mathbf{A}$ may be composed of $k*p$ rows and $k$ columns; and $(iii)$ multiply $\mathbf{A}$ with $(v_1, v_2, \ldots, v_k)$ to 
generate a vector of $\mathbf{c} = (c_1, c_2, \cdots, c_{k*p})$ of $k*p$ elements. The multiplication 
is formally illustrated below: 
\begin{equation} 
    \begin{pmatrix}
        a_{1,1} & a_{1,2} & \ldots & a_{1,k} \\
        a_{2,1} & a_{2,2} & \ldots & a_{2,k} \\
        \vdots & \vdots & \ldots & \vdots \\
        a_{n,1} & a_{n,2} & \ldots & a_{k*p,k}
    \end{pmatrix}
    \begin{pmatrix}
        v_{1} \\
        v_{2} \\
        \vdots \\
        v_{k}
    \end{pmatrix}
    =
    \begin{pmatrix}
        c_1 \\
        c_2 \\
        \vdots \\
        c_{k*p}
    \end{pmatrix}
\end{equation}
A node $v$ communicates 
%in the network sending 
the coded shards $c_1, c_2, \cdots c_{k*p}$ (together with the coefficients that
generated them) to its neighbors.
%in $N_{meta}(v)$. 
% NN: the following is mentioned in the algorithm.
% Therefore, a message exchanged between peers consists of the following  $\tup{id, \mathbb{F}_q, k, c}$, where $id\in\msgSet$ is the identifier of the published message, 
% $k \in \mathbb{N}$, and $c \in (\mathbb{F}_q)^{k}$.  
The matrix $\mathbf{A}$, being randomly generated, can be shown to be invertible with sufficiently high probability.  In the event that a random matrix is generated that is not invertible, one can discard it and wait to receive more shards until an invertible matrix is generated.
}

%\remove{
\myparagraph{{\bf Cryptographic tools:}} We assume that there is an authentication scheme in place that supports two operations: \act{sign}() and \act{verify}(). A sender peer $\pr\in\idSet$ may use \act{sign}($\pr{}, m$), given 
a message $m$ and his identifier $\pr{}$, to generate a signature $s$ for $m$. 
Given $s$, a receiver peer may use \act{verify}($\pr{},m,s$) that evaluates 
to true iff $\pr{}$ executed \act{sign}($\pr{},m$) in some previous step. 
We assume that signatures are \emph{unforgeable}, i.e. no process (including the Byzantine ones) other than $\pr{}$ may invoke \act{sign}($\pr{}, m$). 

We also assume the use of \emph{collision-resistant} cryptographic hash function
$\hash{}: \{0, 1\}^* \rightarrow \{0, 1\}^n$, which on sufficiently large output size $n$, and inputs $x$, $y$ the following properties hold: 
%\begin{itemize}
%\item[(i)] 
(i) {\em Deterministic}: if $x=y$, then $\hash{}(x)=\hash{}(y)$; 
%The function produces the same output for the same input every time.
%\item[(ii)] 
(ii) {\em Fixed Output Length}: The output has a fixed size $n$, regardless of input size; and 
%\item[(iii)] 
(iii) {\em Collision Resistance}: if $x\neq y$, then $\hash{}(x)\neq \hash{}(y)$ with very high probability.
%Finding any two distinct inputs that produce the same hash value should be infeasible in practice.
%\end{itemize}
%}

%\remove{
\myparagraph{\bf Adversarial Model:} We assume that a subset of peers may be Byzantine.
To prevent impersonation attacks, we required every message exchanged between peers to be signed 
with a valid signature, which can only be generated by the sender peer.  
Byzantine peers can, however, produce and propagate a \textit{faulty} shard $c_f$, performing what we call a \textbf{\textit{pollution attack}}, and preventing other peers from decoding 
the message contents. We say that shard $c_f$ is faulty (or \textit{polluted}) if: (i) the coefficients in $c_f$ are altered, (ii) the encoded data in $c_f$ are corrupted, or (iii) $c_f$ is not a 
valid linear combination of other shards. The pollution may spread in the 
network when non-Byzantine peers may attempt to produce shards by linearly combining 
correct with polluted shards.
%}

\section{libp2p Gossip Protocol: \gossipsub{}}
\label{sec:libp2p}
In practice, \optpp{} aims to replace the gossip protocol in \emph{libp2p},
a well established and documented peer network stack \cite{libp2p}. 
Thus, before proceeding with the description of the \optpp{} algorithm,
in this section we examine the details of \gossipsub{}, the current 
gossip algorithm that is currently used within libp2p. 
%
%\paragraph{\bf{Galois Gossiping in libp2p.}}
\emph{libp2p} provides two main gossiping protocols: {\bf \emph{Floodsub}} and {\bf \emph{Gossipsub}}~\cite{libp2p-gossipsub-spec, libp2p-pubsub}.
\begin{itemize}
    \item {\bf \emph{Floodsub}} is a simple approach where each node forwards every message it receives to all its peers, ensuring broad dissemination with some degree of message redundancy.
    \item {\bf \emph{Gossipsub}} provides an improvement on Floodsub in the following ways: (i) by organizing messages into topics, with nodes only forwarding messages to peers subscribed to those topics, (ii) by defining 
    a network degree $D$ to limit the number of peers each node exchanges
    full messages with, and (iii) by utilizing small control messages to 
    optimize the network traffic.    
    % This reduces redundancy and enhances efficiency. Additionally, Gossipsub introduces mechanisms to control the number of message forwards, by way of several control messages, further optimizing network traffic.
\end{itemize}

\begin{figure}[ht]
    \centering
    \includegraphics[width=0.45\textwidth]{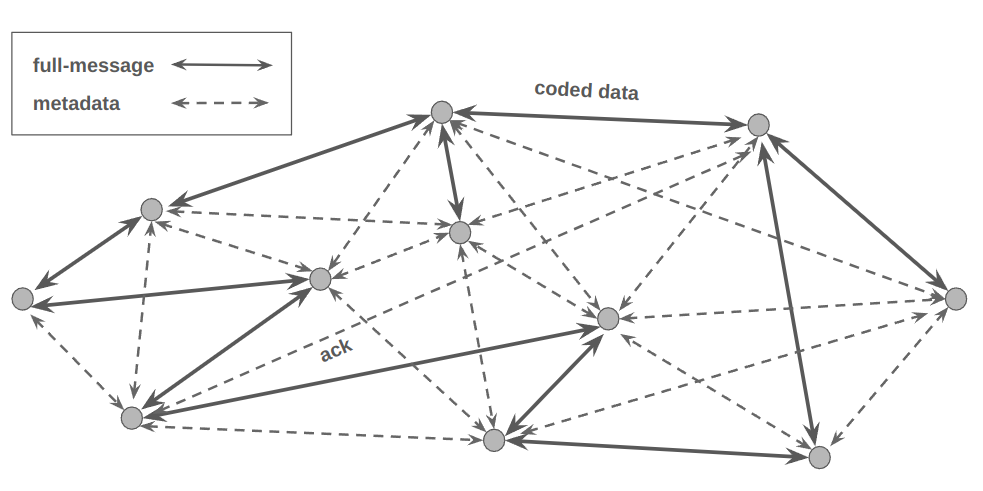}
    \caption{
The figure illustrates two distinct types of peers in the libp2p network: full-message (or mesh) peers and metadata peers} 
% which facilitate bidirectional data communication. Full-message peers operate within a sparsely connected network to transmit substantial payloads, while metadata peers form a densely connected network primarily for exchanging control messages.}
    \label{fig:communication-graph}
\end{figure}

\myparagraph{\bf \gossipsub{} Network Overlay.}
%The following network description is based on the libp2p documentation. 
\nn{In Gossipsub, peers establish connections through two types of peerings: {\bf \emph{full-message peerings}} and {\bf \emph{metadata-only peerings}}. These two types of connections define the network graph (see Fig. \ref{fig:communication-graph}).
\textbf{Full-message} (or \textit{mesh}) peers operate within a sparsely connected network with each peer connected to a degree $D$ other peers, to transmit entire messages across the network. \textbf{Metadata} (or \textit{connected}) peers form a densely connected network primarily for exchanging control messages.}
\remove{
%\begin{itemize}
    %\item 
    \noindent{\bf \emph{Full-message peerings}} enable the transmission of entire messages across the network. This portion of the network is intentionally sparse, with each peer connected to only a limited number of others, i.e. the degree $D$. This sparse network is referred to as the mesh in the Gossipsub specification, and the peers within it are known as mesh members.
    %\item 
    \noindent{\bf \emph{Metadata-only peerings}} involve connections where peers only exchange metadata about available messages, not the full content. This network is more densely connected, ensuring that all peers are informed about the existence of messages and the status of the mesh, without the overhead of transmitting the full message content.
%\end{itemize}
}

The rationale for limiting full-message peerings is to reduce network traffic and increase the available bandwidth. 
% In Gossipsub, each peer only forwards messages to a limited subset of peers, rather than broadcasting to every peer in the network. This approach reduces unnecessary bandwidth consumption.
%
In libp2p's default Gossipsub implementation, $D=6$  with an acceptable range of 4 to 12. The network degree strikes a balance between several key factors: speed, reliability, resilience, and efficiency. Higher $D$
%peering degree 
improves network coverage, and thus message delivery speed, reliability by ensuring messages reach all subscribers, and fault-tolerance by reducing the impact of any peer disconnections. However, increasing the degree raises bandwidth demands and network congestion, as redundant copies of each message
are generated.

\myparagraph{\bf Control Messages.}
By controlling the number of full-message peerings, the network can optimize for both performance and resource efficiency.  The gossipsub protocol leverages several control messages in order to manage the topology of the gossip network and its peer-to-peer connections (see \cite{gossipsub-specification}).  
\remove{
The control message following largely mirrors \cite{gossipsub-specification}.

\begin{itemize}
    \item \textbf{GRAFT} A node that wishes to signal to its peers that it is ready to gossip to one or several topics will send this message.  A peer receiving this message indicates that the peer has been added to the node's local topology for a particular topic.
    \item \textbf{PRUNE} A node that no longer wishes to participate in a topic will send a message to its peers indicating that those peers have been removed from the local mesh topology for a particular topic.
    \item \textbf{IHAVE} A node can gossip a collection of messages that it has recently seen to peers by sending this type of message, alerting them to how listening on the topics containing those messages could be helpful.
    \item \textbf{IWANT} Having seen the message identifiers announced by peers via \textbf{IHAVE}, a node can send this message to request the contents of the identifiers that it names.
    \item \textbf{IDONTWANT}  As soon as a node has received a message, it will immediately broadcast this message to its peers, who, in turn, will carry a list of message ids to avoid re-sending to those who have already seen it.
\end{itemize}
}

\section{OptimumP2P: The Galois Gossip Protocol}
\label{sec:optp2p:gossip}

In this section we present the \optpp{} Galois Gossip protocol that extends ideas 
included in the \gossipsub{} protocol included in libp2p. 

\begin{table}[h!]
\centering
\begin{tabular}{|l|l|}
\hline
\textbf{Parameter} & \textbf{Description} \\ \hline
$N(v)$ & Set of \textit{connected} neighbors of $v$ in $G=(V,E)$ \\ \hline
%$N_{meta}(v)$ & Set of metadata neighbors of $v$ in $G$ \\ \hline
$N_{mesh}(v)$ & Set of full-message (\textit{mesh}) neighbors of $v$ in $G$ \\ \hline
%$t_{req}$ & The maximum time to wait before a request is sent to neighbors \\ \hline
%$t_{delete}$ & The maximum retention time of an $id$ in the $\text{codedBuffer}$ \\ \hline
$t_{heartbeat}$ & The heartbeat time for relaying {\sc ihave} messages \\ \hline
$k$ & Rank of coded-stripes of a message \\ \hline
$r$ & The forwarding threshold \\ \hline
$p$ & Published shard multiplier \\ \hline
%$D$ & Number of full-message neighbors, i.e., $|N_{full}(\pr)|$ \\ \hline
%$f$ & Fraction of $k$-dimensions covered by the coded-stripes of a specific message in a node \\ \hline
\end{tabular}
\caption{\bf{Parameters encoded in the protocol for each node $v$}}
\label{tab:parameters}
\end{table}

% \optpp{} Galois Gossip 
% %is a simple algorithm that 
% relies on three core mechanisms: (i) utilizes RLNC for spliting 
% published messages into shreds, (ii) it uses a peering degree $D$ as defined in \gossipsub{} for defining an overlay network and controlling the bandwidth utilization, and (iii) 
% %a single 
% uses control messages to control information diffusion. Table~\ref{tab:parameters} presents the fix parameters known to each node in the service. 

\begin{algorithm*}[!ht]
\begin{algorithmic}[2]
    \begin{multicols}{2}
        {\scriptsize
            \Statex
            \Part{Data Types}{
                \State $\valSet$: set of allowed message values
                \State $\mathcal{M}\subseteq H$: set message identifiers 
                \State $ \mathfrak{S}$: set of signatures
                %\State $\idSet$: set of node identifiers
            }\EndPart
            
            \Statex
            
            \Part{Parameters}{ \label{line:parameters}
                %\State $\mathcal{T}$: set of topic identifiers
                \State $N(\pr)\subseteq \vertexSet$: set of neighbors of $\pr\in\vertexSet$ in $G=(V,E)$
                %\State $N_{meta}(v)\subseteq N(v)$: set of metadata neighbors of $\pr$
                \State $N_{full}(v)\subseteq N(v)$: set of full-message neighbors of $\pr$ 
                \State $t_{heartbeat}$: time interval for garbage collection
                \State $k\in \Nat$: fragmentation parameter
                \State $r\in \Nat$: forwarding threshold
            }\EndPart

            \Statex

            \Part{State Variables}{ \label{line:rugby:state}
                %\State $peers\subseteq \idSet$ init $\{i\}$
                %\State $\act{shard}\in \idSet\times\mathbb{Z}_{\geq 0}\times\mathbb{F}^k\times\mathcal{M}$ init $\bot$
                %\State $msgType \in \{${\sc ihave}, {\sc iwant}, {\sc idontwant}$\}$
               \State $\act{msgBuffer}\subseteq\mathcal{M}\times \valSet$, published message buffer  initially $\emptyset$ 
               \State $\act{sendBuffer}[m]\subset N(\pr)^2 \times\mathbb{Z}^+_{\geq 0}\times(\mathbb{F}_{2^8})^k$,  send buffer initially $\emptyset$ 
               % $\act{rcvBuffer}[m]\subset N(\pr)\times $, packets received buffer initially $\emptyset$ 
                \State $\act{shardSet}[m] \subseteq N(\pr)^2\times\mathbb{Z}^+_{\geq 0}\times(\mathbb{F}_{2^8})^k\times \mathfrak{S}$ shards for  $m\in\mathcal{M}$ init $\emptyset$
                \State $\act{msgDecoded}[m] \in \valSet \times 2^{\act{shardSet}[m]}$, \WRP decoded value and shards for $m\in\mathcal{M}$ initially $\bot$
                %\State $\act{codeBuffer}$: $map[id]: coded~elements$, initially empty
                %\State $\act{codeBufferTimes}$: $map[id]: Time$, initially $\bot$
                % \State $\act{hasBasisInNbrs}$: $map[v]:[map[id]: bool]$, $v \in N(v)$, 
                % \Statex ~~~~~~~initially empty
                \State $\act{isDone}[m]\subseteq{N(v)}$, set of peers decoded $m\in \mathcal{M}$ init $\emptyset$ 
                %$map[v]:[map[id]: bool]$, $v \in N(v)$, 
                %\Statex ~~~~~~~initially empty
                
                \State $\act{doneSent}[m]\subseteq{N(v)}$, \WRP peers we have sent \mtype{idontwant} for  $m\in \mathcal{M}$ init $\emptyset$ 
                %\State $\act{garbageCollect}\subseteq \mathcal{M}$ initially $\emptyset$
                \State $\act{iWant}[m]\in {N(v)}$, peer to send \mtype{iwant} for $m\in \mathcal{M}$ init $\bot$ 
                {\color{olive}
                %\State $\act{msgSign}[m]$, message signature, initially $\emptyset$
                \State $\act{malShards}[m]\subseteq N(\pr)^2\times\mathbb{Z}^+_{\geq 0}\times(\mathbb{F}_{2^8})^k\times \mathfrak{S}$, initially $\emptyset$
                \State $\act{malPeers}[m]\subseteq {N(v)}$, initially $\emptyset$
                \State $\act{quaPeers}[m]\subseteq  {N(v)}$, initially $\emptyset$
                \State $\act{quaShards}[m]\subseteq N(\pr)^2\times\mathbb{Z}^+_{\geq 0}\times(\mathbb{F}_{2^8})^k\times \mathfrak{S}$, initially $\emptyset$
                \State $\act{isPolluted}[m]\in\{true,false\}$, initially $false$
                }
            }\EndPart

            \Statex
            
            \Part{Signature}{ \label{line:operations}
                \State {\bf Input:}
                \State\T $\act{publish}(\mathbf{v})$, ~$\mathbf{v}\in \valSet$
				\State\T $\act{receive-shard}(m, s)$, ~$m\in \mathcal{M}$, $s\in N(\pr)\times\mathbb{Z}^+_{\geq 0}\times(\mathbb{F}_{2^8})^k$
                \State\T $\act{receive-done}(m, \mtype{idontwant})$, ~$m \in \mathcal{M}$
                \State\T $\act{receive-ihave}(m, \mtype{ihave})$, ~$m \in \mathcal{M}$
                \State\T $\act{receive-iwant}(m, \mtype{iwant})$, ~$m \in \mathcal{M}$
				
			 	\State {\bf Output:}
				\State\T $\act{deliver}(\mathbf{v})$, ~$\mathbf{v}\in \valSet$
				\State\T $\act{send-shard}(m, s)$, ~$m\in \mathcal{M}$, $s\in N(\pr)\times\mathbb{Z}^+_{\geq 0}\times(\mathbb{F}_{2^8})^k$
                \State\T $\act{send-done}(m, \mtype{idontwant})$, ~$m \in \mathcal{M}$
                \State\T $\act{send-ihave}(m, \mtype{ihave})$, ~$m \in \mathcal{M}$
                \State\T $\act{send-iwant}(m, \mtype{iwant})$, ~$m \in \mathcal{M}$
	 
			 	\State {\bf Internal:}
                %\State\T $\act{process-shard}(s, pr)$, ~$s \in Shards$, $pr \in \nodeSet$
                \State\T $\act{generate-shards}()$
                \State\T $\act{decode-message}(m)$, $m\in\mathcal{M}$
            }\EndPart
        }
    \end{multicols}
    \end{algorithmic}
\caption{\optpp{} Gossip: Data Types, Parameters, State and Signature at node $\pr$}
\label{algo:optp2p:signature}
\end{algorithm*}

The \optpp{} protocol primarily relies on a push-based system, in which 
the publisher of a message pushes shards to its peers, who then forward these shards 
to their own peers. The protocol has a fallback for when this system doesn't work;
nodes that don't get enough shards to decode a message can request more shards
from their peers as necessary. \optpp{} uses a peering degree $D$ as defined 
in \gossipsub{} for defining an overlay network. 
%and controlling the bandwidth utilization.

At a high level, \optpp{} works in four stages:
%\begin{enumerate}
    %\item[]
    
    \noindent{\bf Publisher Propagation:} Once a peer $\pr$ receives a $\act{publish}(\mathbf{v})$, it first divides 
    the message to be published into $k$ fragments, and encodes those fragments using RLNC into $p*k$ \textit{shards}. It then sends these shards to its full-message neighbors.
    
    %\item[]
    \noindent{\bf Shard Processing:} When a node $\pr'$ receives a shard, it adds it to its set of shards for the message. If it has enough shards to decode the message, it does so.
    
    %\item[]
    \noindent{\bf Shard Forwarding:} After having processed an incoming shard, a node conditionally forwards it to its own peers. The conditions depend on: (i) 
    whom the shard came from, (ii) how many shards the node has locally, and (iii) the status of the node's peers.
    
    %\item[]
    \noindent{\bf Requesting Additional Shards:} Periodically, a peer that does not have enough shards to decode requests more shards from its peers. Peers that receive such a request supply with more shards. 
%\end{enumerate}

% In high level, the algorithm 
% can be described by the following steps:
% \begin{enumerate}
%     \item[]{\bf Initial Encoding:} Once a peer $\pr$ receives a $\act{publish}(m)$ request it encodes using RLNC
%     the contents of $m$ into $k$ shreds.
%     \item[]{\bf Publisher Propagation:} The publisher $\pr$ then sends, either one of the $k$ initial shreds or a linear combination of 
%     those, to each $D$ of its full-message neighbors.
%     \item[]{\bf Peer Recoding and Forwarding:} when a peer $\pr'$ receives a shred for $m$ it forwards that shred or a linear 
%     combination with other shreds for $m$ to its neighbors
%     \item[]{\bf Peer Decoding and Delivery:} when a peer $\pr'$ receives enough shreds to decode the contents of $m$ it 
%     sends a message $\tup{m,\text{{\sc iamdone}}}$ to \textit{all} of its neighbors and invokes $\act{deliver}(m)$ 
%     \item[]{\bf Peer Control:} a peer $\pr'$ stops forwarding shreds for $m$ if it received the {\sc imadone} message from all of its $D$ full-message neighbors or a timeout for $m$ has expired.
% \end{enumerate}
\subsection{Optimizations}
To further improve performance, \optpp{} adopts four minor optimizations (color coded in Algorithm~\ref{algo:optp2p}): 
(i) publisher flooding ({\color{magenta} magenta}), (ii) forwarding threshold ({\color{cyan} cyan}), 
(iii) control messages ({\color{purple} red}).

\myparagraph{Publisher Flooding:} The algorithm is designed to aggressively forward shards created by the publisher, 
since these shards are created early in the message's lifecycle, and always carry
new degrees of freedom. Thus, 
the publisher sends shards to all of its neighbors, i.e. $N(\pr)$ and not only to mesh peers, i.e. $N_{mesh}(\pr)$,
aiming 
%This action aims 
to expedite the dissemination of the degrees of freedom. 

\myparagraph{Forwarding Threshold:} Each non-publisher node for a message $m$, maintains a forwarding threshold $r$ and creates and forwards a new shard whenever it collects more than $\frac{r}{k}$ shards in its local set for $m$. This shard is then sent to the node's peers in $N_{mesh}(\pr)$. This in contrast to the publisher's policy, as we aim to reduce 
unnecessary propagation
of shards that are unlikely to carry new degrees of freedom to a node's peers.

\myparagraph{Control Messages:} We use control message similar to the \gossipsub{} protocol 
to suppress unnecessary message transmissions and facilitate dissemination of shards to isolated nodes in case of network partitions. The control messages used are: {\sc idontwant, ihave, iwant}.

% \paragraph{\bf{Parameters encoded in the protocol for each $v$}}
% \begin{enumerate}
%     \item $N(v)$: set of neighbors in $G=(V,E)$
%             \item $N_{meta}(v)$: set of metadata neighbors of $v$ in $G$
%             \item $N_{full}(v)$: set of full-message neighbors in $v$ in $G$
%     \item $t_{req}$: the maximum time to wait before a request is sent to neighbors
%     \item $t_{delete}$: the maximum retention time of an $id$ in the $\act{codedBuffer}$
%     \item $k$: rank of coded-stripes of a message
%         \item $f$: fraction of $k$-dimensions covered by the coded-stripes of a specific message in a node.
% \end{enumerate}

\begin{algorithm*}[!ht]
\begin{algorithmic}[2]
    \begin{multicols}{2}
        {\scriptsize
           
            \Part{Transitions}{ \label{line:transitions}

            \State\Comment{message publishing}
    		\Input{$\act{publish}(\mathbf{v})_{v}$\label{line:publish-start}}
    		{%Eff
              \State{$\act{msgBuffer}\gets \act{msgBuffer} \cup\{\tup{\hash{}(\mathbf{v}), \mathbf{v}}\}$}
    		}\EndInput

            \Statex

            \State\Comment{message delivery}
    		\Output{$\act{deliver}(\tup{m, \mathbf{v}})_{v}$ \label{line:deliver-start}}
    		{%Pre
                 \State $\act{msgDecoded}[m]\neq \bot$
            }
            {%Eff
                \State $\tup{\mathbf{v}_{dec},*}\gets \act{msgDecoded}[m]$
                \State $\tup{m,\mathbf{v}}\gets\tup{m,\mathbf{v}_{dec}}$
                %\State \textbf{return} $\tup{id, \mathbf{v}}$
    		}\EndOutput
    		%

      %       \State\Comment{subscribe invocation}
    		% \Input{$\act{subscribe}(t)_{v}$\label{line:subscribe-start}}
    		% {%Eff
      %         \State{$\act{subTopics}\gets \act{subTopics} \cup\{\t\}$}
    		% }\EndInput
    		
    		\Statex

        \State\Comment{Generate and encode data}
            \Internal{$\act{generate-shards}()_{\pr}$}
            {%Pre
                \State $\tup{m, \mathbf{v}} \in \act{msgBuffer}$
                % \State $\act{shardSet}_m \gets \{\tup{s, c_s}: \tup{m,\pr,s,c_s}\in \act{shardSet}\}$
                % \State $\act{shardSet}_m = \emptyset$
            }
            {%Eff
                \State $S\gets\act{RLNCencode}(\mathbf{v}, k*|
                N_{mesh}(\pr)|)$
                %\State $\act{shardSet}[m]\gets \{(\pr, \tup{\pr, s, c_s}, \act{sign}(\pr, \tup{\pr, s, c_s})): \tup{s, c_s}\in S \}$
                \State {\color{magenta} $\act{sendBuffer}[m]\gets\{\tup{v',\tup{\pr, s, c_s}}: v'\in N(\pr)~\wedge~\tup{s, c_s}\in S\}$}
                % \State\Comment{Decompose original message into basic stripes}
                % \State $\mathbf{c}_{uncoded} \gets v_{1} + ... +v_{k}$ 
                % \State
                % \Comment{generate a square matrix of size $k$ of field $\mathbb{F}_q$}
                % \State $\mathbf{A}\gets\act{generate-random-matrix}(k)$
    
                % \State $\mathbf{c} \gets \mathbf{A}\mathbf{v}$
                
                % %\State $\act{codeBuffer}[id] \gets \{\}$
                % \State\Comment{add the uncoded and coded stripes to $\act{codeBuffer}$}
                % \For {$i\in {1, \cdots, k}$}
                %     \State $\act{codeBuffer}[id] \gets \act{codeBuffer}[id] \bigcup_{i=1}^{k} v_{i}$ 
                %     \State $\act{codeBuffer}[id] \gets \act{codeBuffer}[id] \bigcup \{ \tup{id, \mathbb{F}_q, k, c_i}_{\pr}\}$ 
                % \EndFor
                % \For{$v'\in N_{full}(v)$}
                %     \State $\act{isDone}[v'][id] \gets \act{false} $ 
                % \EndFor
                % \State\Comment{record the time of adding the message} 
                % \State $\act{codeBufferTimes}[id] \gets time.Now()$
                %\State \Comment{mark all neighbors as not done}
                %\State $\act{isDone}[m] \gets \emptyset $
                %\State $status\gets propagate$
                % \State\Comment{store a hash of the content}
                %\State { \color{olive} $\act{msgSign}[m]\gets\tup{\pr, \act{sign}(\pr, m)}$}
            }\EndInternal

            %\Statex

    		% \State \Comment{If $v$ is a publisher sends to all}
    		% {\color{magenta}
            
      %       \Output{$\act{send-shard}(\tup{m', \tup{p',s', c_s'}})_{v,v'}$\label{line:send-encoded-data-start}}
      %       {%Pre
      %           \State $v'\in N(v)$
    		% 	\State $v'\notin \act{isDone}[m]$ \Comment{do not send to peer decoded $m$}
      %           \State $\tup{v', \tup{p, s, c_s}}\in \act{sendBuffer[m]}$
      %           \State $v = p$
      %    	}
    		% {%Eff
    		%     \State $\tup{m', \tup{p',s', c_s'}}\gets \tup{m, \tup{p,s, c_s}}$
      %           \State $\act{sendBuffer}[m]\gets \act{sendBuffer}[m]\setminus \{\tup{p, s, c_s}\}$
      %       }\EndOutput
      %       }

            \Statex
 
      %       \State \Comment{If $v'$ is not done yet send encoded data to $v'$}
    		% \Output{$\act{send-shard}(\tup{id, \mathbb{F}_q, k, c}_{\pr}, { \color{magenta} hid})_{v,v'}$\label{line:send-encoded-data-start}}
    		\State \Comment{If $v'$ is not done yet send encoded data to $v'$}
    		\Output{$\act{send-shard}(\tup{m', \tup{p',s', c_s'}, sig})_{v,v'}$\label{line:send-encoded-data-start}}
            {%Pre
                %\State $v'\in N_{full}(v)$
    			\State $\tup{v', \tup{p, s, c_s}}\in \act{sendBuffer[m]}$
                \State $v'\notin \act{isDone}[m]$ \Comment{do not send to peer decoded $m$}
                \State $v' \neq p$\Comment{do not send to the publisher}
                \State { \color{olive} $\act{isPolluted}[m] = false$}
                %\State $status=propagate$
            %\State $wCounter\gets wCounter+1$
    		}
    		{%Eff
    		    \State $\tup{m', \tup{p',s', c_s'}}\gets \tup{m, \tup{p,s, c_s}}$
                \State $\act{sendBuffer}[m]\gets \act{sendBuffer}[m]\setminus \{\tup{v', \tup{p, s, c_s}}\}$
                \State { \color{olive} $sig \gets \act{sign}(\pr, \tup{p, s, c_s})$}     
    		}\EndOutput
                
            \Statex

            \State\Comment{Receive coded shards from v'}
            \Input{$\act{receive-shard}(\tup{m, \tup{p, s, c_s}, sig})_{\pr', \pr}$\label{line:receive-message-start}}
            {%Eff
                %\If{{\color{magenta} $\pr'\notin \act{malPeers}[id]\cup\act{quaPeers}[id]~\wedge~\act{valid}(h)$}}
                % \State\Comment{Recode here before storing}
                % \State$\mathbf{A} \gets \act{generate-random-matrix}(k)$
                % \State$\mathbf{c}_{recoded} \gets \mathbf{A}c$
                \If{$msgDecoded[m] = \bot~\wedge~{\color{olive} \act{verify}(\pr',\tup{p, s, c_s},sig)}$}
                    \If{${\color{olive} \pr'\notin \act{malPeers}[m]\cup\act{quaPeers}[m]}$}
                        \State $\act{shardSet}[m]\gets\act{shardSet}[m]\cup\{(v', \tup{p, s, c_s}, {\color{olive} sig})\}$ 
                        \If{{\color{cyan} $|\act{shardSet}[m]| > \frac{r}{k}$}$~\wedge~{\color{olive} \act{isPolluted}[m] = false}$}
                            \State $S\gets \{s: (*,s,*)\in \act{shardSet[m]}\}$
                            \State $\tup{s',c_{s'}}\gets \act{RLNCrecode}(S)$
                            \State $B\gets \{\tup{v',\tup{p, s',c_{s'}}}: v'\in N_{mesh}(\pr)\setminus\{p\}\}$
                            \State $\act{sendBuffer}[m]\gets \act{sendBuffer}[m]\cup B$
                        \EndIf
                    \Else
                        \State\Comment{accept only if coming from quarantine neighbor}
                        \If{${\color{olive} \pr'\in\act{quaPeers}[m]}$}
                        \State {\color{olive} $\act{quaShards}[m]\gets\act{quaShards}[m]\cup\{(v',\tup{p, s, c_s},sig)\}$ }
                        \EndIf
                    \EndIf
                \EndIf
                %\If{ {\color{magenta} $\act{valid}(h)$} }
                %\State {\color{magenta} $msgHash[id]\gets h$ }
                % \Else
                % \If{{\color{magenta} $\pr'\in \act{quaPeers}[id]~\wedge~\act{valid}(h)$}}
                % \State {\color{magenta} $\act{checkCodes}[id]\gets\act{checkCodes}[id]\cup\{\tup{id, \mathbb{F}_q, k, c}_{\pr}\}$ }
                % \State {\color{magenta} $msgHash[id]\gets h$ }
                % \EndIf
                % \EndIf
            }\EndInput

            \Statex

            \State\Comment{check whether a message is decodable}
            \Internal{$\act{decode-msg}(m)_{\pr}$}
            {%Pre
                %\State $|\act{codeBuffer}[id]|>k$
                \State $\exists S\subseteq \{s: (*,s,*)\in\act{shardSet}[m]\}$ s.t. $|S|=k~$
                %{\color{magenta} $\wedge~\hash{}(\act{decode}(S))= msgHash[id]$}
                \State $\mathbf{v}_{dec}\gets \act{RLNCdecode}(S)$
                \State $m = \hash{}(\mathbf{v}_{dec})$
            }
            {%Eff
                % \State \Comment{random subset of at least $k$ shreds} 
                % \State $S\subseteq \act{codeBuffer}[id]$ s.t. $|S|\geq k$
                % \State $\mathbf{v}_{dec}\gets \act{decode}(\act{codeBuffer}[id])$
                \State $\act{msgDecoded}[m]\gets \tup{\mathbf{v}_{dec},S}$
                \State {\color{olive} $\act{isPolluted[m]}\gets false$}
            }
            \EndInternal

            \Statex
            {\color{purple}
            \State\Comment{when we decode, send done to mesh neighbors}
            \Output{$\act{send-done}(\tup{m, \mtype{idontwant}})_{v,v'}$\label{line:send-done-start}}
            {%Pre
                \State $\act{msgDecoded}[m]\neq \bot$ 
                \State $v'\in N_{mesh}(v)$
    			%\State $\act{codeBuffer}[id]\neq\emptyset$
                \State $v'\notin\act{doneSent}[m]$
            %\State $wCounter\gets wCounter+1$
    		}
    		{%Eff
    			\State $\act{doneSent}[m] \gets \act{doneSent}[m]\cup\{v'\}$
    		}
            \EndOutput

            \Statex
            
            \State\Comment{Receive done message from v'}
            \Input{$\act{receive-done}(\tup{m, {\mtype{idontwant}}})_{v',v}$}
            {%Eff
                 \State $\act{isDone}[m] \gets \act{isDone}[m]\cup\{v'\} $
                 % \State\Comment{if all neighbors are done mark the id for removal}
                 % \If{$N_{full}(v)\subseteq\act{isDone}[m]$}
                 %    \State $\act{garbageCollect}\gets\act{garbageCollect}\cup\{m\}$
                 % \EndIf
            }
            \EndInput

            \Statex 
            
            \State\Comment{when we decode, send ihave to mesh neighbors}
            \Output{$\act{send-ihave}(\tup{m, \mtype{ihave}})_{v,v'}$\label{line:send-done-start}}
    		{%Pre
                \State $time.Now()-heartbeat> t_{heartbeat}$
                \State $\act{msgDecoded}[m]\neq \bot$ 
                \State $v'\in N(v)$
    			%\State $\act{codeBuffer}[id]\neq\emptyset$
                \State $v'\notin\act{isDone}[m]$
            %\State $wCounter\gets wCounter+1$
    		}
    		{%Eff
    			\State $heartbeat \gets time.Now()$
    		}\EndOutput

            \Statex
            
            \State\Comment{Receive ihave message from $v'$}
            \Input{$\act{receive-ihave}(\tup{m, {\mtype{ihave}}})_{v',v}$}
            {%Eff
                 \State $\act{isDone}[m] \gets \act{isDone}[m]\cup\{v'\} $
                 \State\Comment{iwant message if we have not decoded $m$}
                 \If{$\act{msgDecoded}[m] = \bot$}
                    \State $iwant[m]\gets v'$
                 \EndIf
                 % \State\Comment{if all neighbors are done mark the id for removal}
                 % \If{$N_{full}(v)\subseteq\act{isDone}[m]$}
                 %    \State $\act{garbageCollect}\gets\act{garbageCollect}\cup\{m\}$
                 % \EndIf
            }
            \EndInput

            \Statex

            \State\Comment{send iwant to $v'$}
            \Output{$\act{send-iwant}(\tup{m', \mtype{iwant}})_{v,v'}$\label{line:send-done-start}}
    		{%Pre
                \State $iwant[m] = v'$
            %\State $wCounter\gets wCounter+1$
    		}
    		{%Eff
                \State $m' \gets m$
    			\State $iwant[m] \gets \bot$
    		}\EndOutput

            \Statex
            
            \State\Comment{Receive iwant message from $v'$}
            \Input{$\act{receive-iwant}(\tup{m, {\mtype{iwant}}})_{v',v}$}
            {%Eff
                 \State\Comment{if we have already decoded $m$}
                 \If{$\act{msgDecoded}[m] \neq \bot$}
                    \State $\tup{*,S}\gets\act{msgDecoded}[m]$
                    \State $\tup{s',c_{s'}}\gets \act{RLNCrecode}(S)$
                    \State $\act{sendBuffer}[m]\gets \act{sendBuffer}[m]\cup\{\tup{v', \tup{p, s',c_{s'}}}\}$
                 \EndIf
                 % \State\Comment{if all neighbors are done mark the id for removal}
                 % \If{$N_{full}(v)\subseteq\act{isDone}[m]$}
                 %    \State $\act{garbageCollect}\gets\act{garbageCollect}\cup\{m\}$
                 % \EndIf
            }
            \EndInput
    }

            % \Statex

            % \State\Comment{Expired message ids are marked for removal}
            % \Internal{$\act{message-timeout}()_{\pr}$}
            % {%Pre
            %     \State $\act{codeBufferTimes}[id]\neq\bot$
            %     \State $(time.Now() -\act{codeBufferTimes}[id]) \geq t_{delete}$
            % }
            % {%Eff
            %     \State \Comment{mark the message for garbage collection}
            %     \State $\act{garbageCollect}\gets\act{garbageCollect}\cup\{id\}$
            % }
            % \EndInternal

            % \Statex

            % \State\Comment{Garbage collect marked messages}
            % \Internal{$\act{garbage-collection}()_{\pr}$}
            % {%Pre
            %     \State $id\in\act{garbageCollect}$
            % }
            % {%Eff
            %     \State \act{Remove} $id$ \act{from} $\act{msgBuffer}$   
            %     \State \act{Remove} $id$ \act{from} $\act{codeBuffer}$   
            %     \State \act{Remove} $id$ \act{from} $\act{codeBufferTimes}$     
            %     %\State \act{Remove} $id$ \act{from} $\act{reqSentAtTimes}$
            %     \State \act{Remove} $id$ \act{from} $\act{doneSent}$
            %     \State \act{Remove} $id$ \act{from} $\act{isDone}$ 
            %     \State \act{Remove} $id$ \act{from} $\act{garbageCollect}$  
            % }
            % \EndInternal    
            }\EndPart%PART TRANSITIONS
        }
    \end{multicols}
    \end{algorithmic}
\caption{\optpp{} Gossip: Transitions at any node $\pr\in\vertexSet$ of graph $G = (\vertexSet, E)$.
%Protocol for gossiping at any node $v, v \in V$ in the graph $G = (V, E)$.
}
\label{algo:optp2p}
\end{algorithm*}

\nn{
\subsection{RLNC External Library}
We assume an external library that 
handles the RLNC encoding, recoding, and decoding process 
and offers the following interface: 

\myparagraph{\act{RLNCencode}($\mathbf{v}, n)$:} The \act{RLNCencode} operation accepts a data value $\mathbf{v}$ 
and uses the RLNC encoding (see Section \ref{sec:model}) to 
generate $n$ shards. 

\myparagraph{\act{RLNCrecode}($S$):} The \act{RLNCencode} 
operation accepts a set of shards $S$ and randomly 
combines the shards in $S$ to generate a new shard $s'$. 
Note that the shards in $S$ may be the result of an 
\act{RLNCencode} or another \act{RLNCrecode} operation. 

\myparagraph{\act{RLNCdecode}($S$):} Last, the \act{RLNCdecode}
operation given a set $S$ of shards, s.t. $|S|\geq k$, 
it attempts to use those shards to decode the original value $\mathbf{v}$.

We use those operations in the specification of \optpp{} 
without providing their detailed implementation as this is 
out of the scope of this work. 
}

\subsection{IOA Specification}
% We now describe the \emph{State Variables} and \emph{Transitions} of the protocol as those appear in Algorithms~\ref{algo:optp2p:signature} and \ref{algo:optp2p}.
The algorithm is formally specified
using the IOA notation \cite{Lynch1996} through group-defined transitions, each characterized by a specific precondition and its effect. We assume that each node operates in a single-threaded mode, executing transitions atomically once their preconditions are met. The execution of these transitions occurs asynchronously. We adopt a \textit{fairness} assumption in the protocol's execution, which suggests that if preconditions are continually met, each node will have infinite opportunities to execute its transitions that satisfy these preconditions. Algorithm \ref{algo:optp2p:signature} presents
the data types, the static parameters, and the state variables used, along with the signature 
of \optpp{}. Algorithm \ref{algo:optp2p} presents the transitions of all the actions in \optpp{}.

% The {\color{magenta} magenta} lines are used during the data pollution mitigation phase and can be 
% ignored for the discussion in this section. We will further discuss that code in Section~\ref{sec:optp2p:rugby}.

% More formally, our protocol is described using the IOA notation through group-defined transitions, each characterized by a specific precondition and its effect. 
% We assume that each node operates in a single-threaded mode, executing transitions atomically once their preconditions are met. The execution of these transitions occurs asynchronously. We adopt a \textit{fairness} assumption in the protocol's execution, which suggests that if preconditions are continually met, each node will have infinite opportunities to execute its transitions that satisfy these preconditions.

% \begin{figure}[h]
%     \centering
%     \includegraphics[width=0.8\textwidth]{buffers.png}
%     \caption{
% The figure shows a node in the Galois gossip protocol, illustrating how it stores sets of coded elements from individual messages and transmits these elements to its neighboring nodes.}
%     \label{fig:buffers}
% \end{figure}

% \begin{figure}
%     \centering
%     \includegraphics[width=0.8\textwidth]{acknowledgement.png}
%     \caption{
% The figure illustrates the two types of control messages a node may send to its neighbors in the context of coded element decoding: one indicating that decoding of a message with a specific identifier is possible, and another signaling that additional coded elements are needed to enable decoding.}
%     \label{fig:acknowledgement}
% \end{figure}

\myparagraph{\bf{State Variables.}} Every node $v$ maintains the following state variables:
%\begin{enumerate}
    
    %\item[] 
    \noindent$\act{msgBuffer}$: a temporary buffer that keeps pairs of message ids and message values that have been requested for publishing. The message id is the hash of the value.
    
    %\item[]  
    \noindent$\act{shardSet}$: a key-value map where the keys are message $id$s and the values are sets of shards for the same message.
    
    %\item[] 
    % \noindent$\act{codeBufferTimes}$: a key-value map where the keys are message $id$s and the values are time instants (based on the local clock at $\pr$) the message $id$ is seen by $\pr$ for the first time.
    
    %\item[] 
    \noindent$\act{msgDecoded}$: a key-value map where the keys are message $id$s and the values are pairs of decoded values with a set of shards that were used during the decoding. 
    % again key-value maps where the keys are nodes $v' \in N(v)$ and the values are a set of vectors or some representation of the basis of a subspace.
    
    %\item[] 
    \noindent$\act{doneSent}$: is a key-value map where the keys are message $id$s and the value is a set of peers to which we have sent the {\sc idontwant} message.
    
    %\item[] 
    \noindent$\act{isDone}$: is a key-value map where the keys are message $id$s and the value is a set of peers from which we have received the {\sc iamdone} message.
    % \item $\act{reqSentAtTime}$: a key-value map where the keys are the message $id$s and the values are time instant (locally in $v$) the last request for more coded stripes was sent to the neighbors $N(v)$ 

    %\item[]
    \noindent$\act{iWant}$: is a key-value map where the keys are message $id$s and the value is the id of a peer from which we want to request {\sc iwant} message.
    
    %\item[] 
    % \noindent$\act{garbageCollect}$: set of message identifiers that indicate which 
    % messages are ready to be garbage collected. 
%\end{enumerate}

% \subsection{External Signature}
% \paragraph{Signature Domain}
% Each message publish should carry a unique identifier $id$ and the $value$ is considered to be a byte buffer or, equivalently, string.
% \begin{itemize}
%     \item publish($id$, $value$)
%     \item subscribe()
%     \item deliver($id$, $value$)
% \end{itemize}

% \begin{figure}[h]
% \begin{tabular}{cc}
%     \includegraphics[width=0.45\textwidth]{buffers.png} & 
%     \includegraphics[width=0.45\textwidth]{acknowledgement.png} \\
% \end{tabular}
% \caption{
% Left: The figure shows a node in the Galois gossip protocol, illustrating how it stores sets of coded elements from individual messages and transmits these elements to its neighboring nodes., Right:
% The figure illustrates the two types of control messages a node may send to its neighbors in the context of coded element decoding: one indicating that decoding of a message with a specific identifier is possible, and another signaling that additional coded elements are needed to enable decoding.}
% \label{fig:acknowledgement}
% \end{figure}

\myparagraph{\bf{Transitions.}} We now describe the input, output and internal actions 
of the protocol. Note that input actions are always enabled and are triggered 
by the environment. Both output and internal actions are executed only if 
their preconditions are satisfied.

\myparagraph{{\bf \act{publish}}:} a node invokes a \act{publish} operation when 
it executes this action. The publish action accepts a value $\mathbf{v}$ to be published and 
generates the message identifier $m$ by hashing the given value. It then  
inserts $\tup{m, \mathbf{v}}$ into a message buffer (\act{msgBuffer}) for
further processing. 

\myparagraph{{\bf \act{deliver}}:} once a message is decoded the \act{deliver} action returns the value of a message to the caller.

\myparagraph{{\bf \act{generate-shards}}:} this action encodes 
%if there are 
pending messages to be published in \act{msgBuffer}
%, this action is triggered to encode each message 
into $p*k$ coded shrards using \act{RLNCencode}. These shards are then stored in the $\act{shardSet}$, and a tuple $\tup{v', s'}$ is added in the \act{sendBuffer} for every peer $v'$ 
of the publisher and every shard $s'$ generated.
%and the time each message was generated (via a location function $time.Now()$) is recorded in the $\act{codeBufferTimes}[id]$.

\myparagraph{{\bf \act{send-shard}}:} is executed to send a shard for a message $m$ to a peer $v'$
%. This action is triggered when corresponding shreds for a message $id$ are available in its 
when an entry $\tup{v', *}\in\act{sendBuffer}[m]$, $v'$ has not yet informed $v$ that has decoded the message $m$,
and $v'$ is not the publisher of $m$. Once the message is sent is removed from the \act{sendBuffer}.
% Note, that the action 
% generates and sends a coded shred resulted from recoding a subset of coded shreds in 
% the $\act{codeBuffer}[id]$. The message $id$, the finite
% field, the size $k$, and the data of the recoded shred are enclosed in each packet sent. 

\myparagraph{{\bf \act{receive-shard}}:} when peer $v$ receives a shard from $v'$ for a message $m$, it 
adds the shard in \act{shardSet} if $m$ is not yet decoded; otherwise it discards the shard. Whenever 
it adds the new shard in its \act{shardSet} and that contains more than $frac{r}{k}$ shards (see optimization (ii),
$\pr$ generates a new shard by \textit{recoding} the shards received 
using the \act{RLNCrecode} action. It then prepares to send the shard generated to all its full peers 
by adding the appropriate entries in the \act{sendBuffer}. Note that this action is a key to the performance 
boost of \optpp{}.
% adds the code shred $\tup{id, \mathbb{F}_q, k, c}$ to $\act{codedBuffer}$ as soon as it is received
% from one of its neighbor. 

\myparagraph{{\bf \act{decode-msg}}:} if a node $\pr$ collects more than $k$ shards in \act{shardSet} 
%in its code buffer 
for a message $m$, then this action is trigger to decode the 
message $m$ and store the outcome along with the shards used for the decoding in 
the $\act{msgDecoded}[m]$ variable.

\myparagraph{{\bf \act{send-done}}:} if a message $m$ is decoded and the receiver 
$\pr'\in N_{mesh}(\pr)$ was not yet informed, i.e. $\pr'$ does not appear in $\act{doneSent}[m]$, 
$\pr$ sends the {\sc idontwant} control message to $\pr'$. 
%Note that this is the only control message sent in \optpp{}.

\myparagraph{{\bf \act{receive-done}}:} upon receiving of a \tup{$m$, {\sc idontwant}} message from $\pr'$, node $\pr$ adds $\pr'$ in its $\act{isDone}[m]$.

\myparagraph{{\bf \act{send-ihave}}:} when a $heartbeat$ timer expires at $\pr$ for a decoded message $m$ 
then $\pr$ sends to $N(\pr)$ peers that are not in \act{isDone[m]} set, an {\sc ihave} message for $m$.

\myparagraph{{\bf \act{receive-ihave}}:} upon receiving of a \tup{$m$, {\sc ihave}} message from a peer $\pr'$, node $\pr$ adds $\pr'$ in its $\act{iWant}[m]$ variable if it have not yet decode $m$ to request shards from $\pr'$.

\myparagraph{{\bf \act{send-iwant}}:} request shards for $m$ from 
$\pr'$.
%send {\sc iwant} message to $\pr'$, to request shards for $m$. 

\myparagraph{{\bf \act{receive-iwant}}:} upon receiving of a \tup{$m$, {\sc iwant}} message from $\pr'$, generate
a new shard by recoding the shards used to decode message $m$. Then queue the new shard to be sent to $\pr'$.

\section{Pollution Avoidance in OptimumP2P: The Rugby Protocol}
\label{sec:optp2p:rugby}
%\subsection{Pollution Mitigation}
Defining pollution as a corruption of shards forwarded in the network, there is a desire to detect and handle that corruption without bringing network gossip to a halt.  We propose the following algorithm in order to mitigate pollution:

% \begin{enumerate}
%     \item After decoding, use a hash (i.e. keccak256) to detect the presence of pollution
%     \item Once pollution is detected, a node initiates a brute-force detection of the culprit(s) by selectively ignoring messages from subsets of peers, making note of which omissions lead to the absence of pollution
%     \item At the end of the brute-force process, a node will penalize the peer score of peers deemed to be culprit(s)
%     \item \texttt{libp2p-peer-scoring} will eventually disconnect peers whose peer score falls below a threshold
%     \item As peers with low reputation are discarded, a design parameter/heuristic should be used to determine when a node should seek additionally well-behaving peers to which it can connect
% \end{enumerate}

\begin{figure*}[h]
    \centering
    \includegraphics[width=0.8\linewidth]{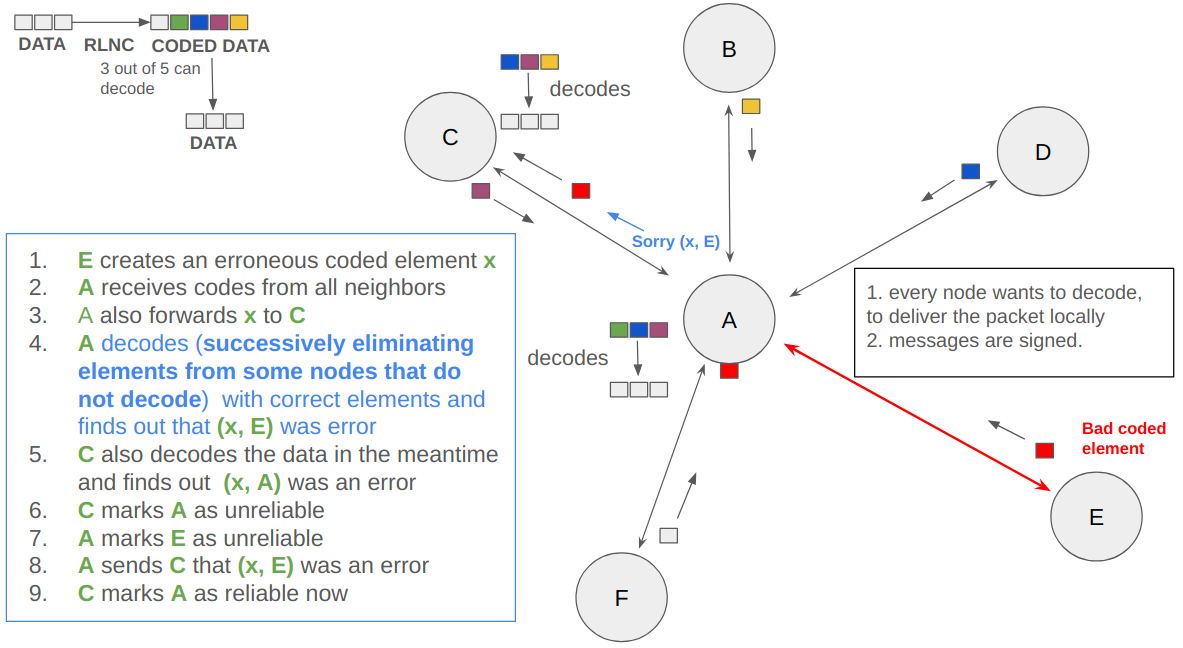}
    \caption{Diagram illustrating the detection of pollution at node A, where the origin of pollution is node E}
    \label{fig:pollution}
\end{figure*}

\myparagraph{\bf Pollution detection:} A node uses the hash of the original message (i.e. keccak256) to detect the presence of pollution. Pollution can be detected while 
    a node attempts to decode a message: (i) either the node has more than $k$ shards
    but fails to decode the message, or (ii) it only succeeds to decode with a proper
    subset of the received shards (i.e., some shards cannot be used to decode the message correctly). 
    % we use a hash (e.g., Keccak256)  to detect after decoding that there was  Byzantine pollution attack. We can use SHA, or we can use the prime field polynomial we discussed. **NB**: unlike homomorphic encryption, there is **no need** for the transmissions to be done over large prime fields because the transmission can be done using a binary extension field, then the data transformed to a prime filed representation **post decoding** in order to compute the polymonial hash, i.e. the field representation for the transmission and that for the hash are entirely decoupled because the hash is computer post decoding.

\myparagraph{\bf Byzantine Peer Identification:} If a node detected a pollution but yet was able to correctly decode the message, then the node initiates a brute-force detection of the polluted neighbors, by selectively ignoring messages from subsets of peers, making note of which omissions lead to the absence of pollution.
    
    % If a peer, say A, finds out that there was a pollution attack, then A checks which of her neighbors (say B, C, or D) was the source of the attack in the following way:

    % She successively removes all of the coded pieces from a neighbor, obtains more coded pieces from other neighbors until she is able to decode. When she is able to decode correctly, she knows which neighbor(s) was (were bad). So, for example A decodes without C, but is still not able to decode with only B and D.  So, she includes C back into the decoding and removes B. Say she is able to decode, she knows now that B was bad. (We may need to go to doubles). 

\myparagraph{\bf Byzantine Avoidance:} Once a node detects that a peer $\pr'$ is polluted, 
    adds $\pr'$ in a list of malicious peers, and stops accepting shards from $\pr'$.\footnote{In case the protocol is integrated in libp2p 
    we can  penalize the \act{libp2p-peer-score} of the polluted peer before we stop accepting messages.}
    % A knows B is bad, she does not accept information from B**, but continues to listen to B for Optimum P2P protocol information.

\myparagraph{\bf Byzantine Finger-Pointing:} A node informs its peers when he/she detects a malicious neighbor $\pr'$, attaching the polluted shard as a proof. Every neighbor 
    receiving this message, verifies that the shard attached indeed cannot be used for decoding the message, and it stops accepting data from $\pr'$. Note that the shard cannot
    be forged by the sender of the alert as each shard is signed by its creator (in this case $\pr'$).
    % ** In our example A told him/her downstream nodes, say E, F and G, that he/she had been polluted. He/She is now able to exhibit to E,F, and G a proof, which is the erroneous coded piece from B, that he/she was not the source of the pollution, but that it was B. If any of E,F and G listened to B, they will also stop accepting data other than protocol data.

\myparagraph{\bf Self Quarantine:} If a node discovers it is the source of pollution (i.e., the Byzantine node), it remains silent until it can decode the message and generate a valid shard. 
    % B is also checking whether he has been polluted.** If he was the Byzantine source of the pollution, then he is unable to blame anyone else and remains isolated. If he has not the source of the pollution then he needs to identify, like A did, the at least one coded piece from at least one node to exonerate him.

\myparagraph{\bf Self-Pollution Alerts:} A node $\pr$ alerts its neighbors as soon as it discovers it has been polluted (even without pointing to the Byzantine source). 
    This allows $\pr$ to maintain its reputation so that it can be eventually reintegrated. Like rugby rules - when you are offside, you raise your hand and get back onside (hence the term \textit{rugby protocol}). Every peer that receives such a self-accused message adds $\pr$ in a quarantine list. 
    
\myparagraph{\bf Pollution Audit:} A node checks every message received from a peer $\pr$ 
    in its quarantine list for decodability. If a message from $\pr$ is a valid shard then $\pr$ is removed from the quarantine. 
%\end{enumerate}

Figure~\ref{fig:pollution} presents an example of the execution of the algorithm where
peer with id $E$ is malicious and its being detected by peer $A$. 

\begin{algorithm*}[!ht]
\begin{algorithmic}[2]
    \begin{multicols}{2}
        {\scriptsize

            \remove{
            \Part{State Variables}{ \label{line:rugby:state}
                \State $\act{msgHash}:map[id]:$message hash, initially empty
                \State $\act{malCodes}:map[id]: 2^{N(v)}\times coded~element$, initially $\emptyset$
                \State $\act{malPeers}:map[id]: 2^{N(v)}$, initially $\emptyset$
                \State $\act{quaPeers}:map[id]:2^{N(v)}$, initially $\emptyset$
                \State $\act{checkCodes}$: set of shreds, initially $\emptyset$
                \State $\act{isPolluted}:map[id]:\{true,false\}$, initially $false$
            }\EndPart
            }

            \Statex
            
           \Part{Signature}{ \label{line:rugby:operations}
                \State {\bf Input:}
				\State\T $\act{receive-alert}(m, \pr_b, \tup{p, s, c_s}, sig)_{\pr',\pr}$, \WRP ~$m\in\msgSet$, ~$s\in\mathbb{Z}^+_{\geq 0}$, $c_s\in(\mathbb{F}_{2^8})^k~$, ~$\pr_b,p,v,v'\in\vertexSet$, ~$sig\in \mathfrak{S}$
                \State\T $\act{receive-polluted}(\tup{m,\text{\sc polluted}})_{\pr',\pr}$, \WRP ~$m\in\msgSet$, ~$\pr_b,v,v'\in\vertexSet$
				
			 	\State {\bf Output:}
				\State\T $\act{send-alert}(m, \tup{\pr_b, s, c_s})_{\pr',\pr}$, \WRP ~$m\in\msgSet$, ~$s\in\mathbb{Z}^+_{\geq 0}$, $c_s\in(\mathbb{F}_{2^8})^k~$, ~$\pr_b,v,v'\in\vertexSet$
                \State\T $\act{send-polluted}(\tup{m,\text{\sc polluted}})_{\pr',\pr}$, \WRP ~$m\in\msgSet$, ~$\pr_b,v,v'\in\vertexSet$
                
			 	\State {\bf Internal:}
			 	\State\T $\act{pollution-discovery}(m)_{\pr}$, ~$m\in\msgSet$, ~$\pr\in\vertexSet$
                \State\T $\act{self-isolate}()_{\pr}$, ~$\pr\in\vertexSet$
                \State\T $\act{check-shards}(m)_{\pr}$, ~$m\in\msgSet$, ~$\pr\in\vertexSet$
            }\EndPart

            \Statex
            
        \Part{Transitions}{ \label{line:rugby:transitions}

            \State\Comment{identify polluted peers}
    		\Internal{$\act{pollution-discovery}(m)_{v}$\label{line:publish-start}}
    		{%Pre
                %\State $|\act{shardSet}[m]| > k$
                \State $\act{msgDecoded}[m] \neq empty$
                \State \Comment{there is pollution}
                \State $\exists S\subset \{s:(*,s,*)\in\act{shardSet}[m]\}$ s.t. \WRP
                $|S|=k~\wedge~\mathbf{v}_{dec}\gets \act{RLNCdecode}(S)~\wedge~\hash{}(\mathbf{v}_{dec})\neq m$
            }
            {%Eff
                \State \Comment{get the set that correctly decoded}
                \State $\tup{*,S_{dec}}\gets \act{msgDecoded}[m]$
                \State\Comment{remove a random element from $S_{dec}$}
                \State $s_r\gets random(s:s\in S_{dec})$
                \State $S_{test} \gets S_{dec}\setminus\{s_r\}$
                \State\Comment{collect polluted shards and their senders}
                \State $S_{pol}\gets \{(\pr_b, s, sig_b): s\in S~\wedge~ (\pr_b,s, sig_b)\in\act{shardSet}[m]$\WRP$~\wedge~\hash{}(\act{RLNCdecode}(S_{test}\cup\{s\})\neq m\}$
                % \State\Comment{collect all nodes sent those shreds}
                % \State$V_{pol}\gets\{\pr': \tup{\pr',*}\in S_{pol}\}$
                \State\Comment{update the list of malicious peers}
                \State $\act{malShards}[m]\gets\act{malShards}[m]\cup S_{pol}$
                \State $\act{malPeers}[m]\gets\act{malPeers}[m]\cup\{\pr_b:(\pr_b, *,*)\in S_{pol}\}$
                \State \Comment{drop polluted shards}
                \State $\act{shardSet}[m]\gets \act{shardSet}[m]\setminus S_{pol}$
    		}\EndInternal

            \Statex
            
            \State\Comment{put your self in quarantine}
    		\Internal{$\act{self-isolate}(id)_{v}$\label{line:self-isolate-start}}
    		{%Pre
                \State $|\act{shardSet}[m]| \geq k$
                \State \Comment{there is pollution}
                \State $\act{msgDecoded}[m] = empty$
                %\State $\nexists S\subseteq \act{codeBuffer}[id]$ s.t. \WRP$|S|=k~\wedge~\hash{}(\act{decode}(S)) = msgHash[id]$
            }
            {%Eff
                \State $\act{isPolluted}[m]\gets true$
    		}\EndInternal

            \Statex
            
            \State\Comment{check shards received from quarantined peers}
    		\Internal{$\act{check-shards}(m)_{\pr}$\label{line:check-codes-start}}
    		{%Pre
                \State $\act{msgDecoded}[m] \neq empty$
                \State \Comment{there is a shard to check}
                \State $(v', \tup{p,s,c_s}, sig)\in \act{checkShards}[m]$
                %\State $\nexists S\subseteq \act{codeBuffer}[id]$ s.t. \WRP$|S|=k~\wedge~\hash{}(\act{decode}(S)) = msgHash[id]$
            }
            {%Eff
                %\State \Comment{get the set that correctly decoded}
                \State $\tup{*,S_{dec}}\gets \act{msgDecoded}[m]$
                \If{$\hash{}(\act{RLNCdecode}(S_{dec}\cup\{\tup{p,s,c_s}\})= m$}
                    \State\Comment{add the good shard in you local buffer}
                    \State $\act{shardSet}[m]\gets\act{shardSet}[m]\cup\{(v',\tup{p,s,c_s}, sig)\}$
                    \State\Comment{reinstated the peer}
                    \State $\act{quaPeers}[m]\gets\act{quaPeers}[m]\setminus\{v'\}$
                \EndIf
                \State $\act{quaShards}[m]\gets\act{quaShards}[m]\setminus\{(v',\tup{p,s,c_s}, sig)\}$
    		}\EndInternal

            \Statex 
            
            \State\Comment{send alert with a proof}
            \Output{$\act{send-alert}(m', \pr'', \tup{p,s,c_s}, sig)_{\pr,\pr'}$ \label{line:send-data-request-start}}
    		{%Pre
                \State $\pr'\in N_{full}(v)$
    			\State $(\pr_b, \tup{p, s_b,c_{s_b}}, sig_b)\in\act{malShards}[m]$
            %\State $wCounter\gets wCounter+1$
    		}
    		{%Eff
                \State $m'\gets m$
                \State $\pr''\gets \pr_b$
                \State $sig\gets sig_b$
                \State $\tup{p,s,c_s} \gets \tup{p, s_b,c_{s_b}}$
    		}\EndOutput

            \Statex
            
            \State\Comment{received a notification of a malicious node}
            \Input{$\act{receive-alert}(m, \pr_b, \tup{p,s,c_s}, sig)_{v',v}$}
            {%Eff
                \If{$\pr_b\notin\act{quaPeers}[m]~\wedge~\act{verify}(\pr_b, \tup{p,s,c_s}, sig)$}
                \State $\tup{*,S_{dec}}\gets \act{msgDecoded}[m]$
                \State $S_t = S_{dec}\setminus random(s: s\in S_{dec})$
                \If{$S_t=\emptyset~\vee~\hash{}(\act{RLNCdecode}(S_{t}\cup\{\tup{p,s,c_s}\})\neq m$}

                %~\wedge~\act{invalid}(s_b)$}
                 % \State $\act{shredToCheck}[id]\gets\act{shredToCheck}[id]\cup \{\tup{\pr_b, \tup{id,*,*,*}_{\pr_b}}\}$
                    \State$\act{malPeers}[m]\gets \act{malPeers}[m]\cup\{\pr_b\}$
                \EndIf
                \EndIf
            }
            \EndInput

            \Statex 
            
            \State\Comment{inform peers that we are polluted}
            \Output{$\act{send-polluted}(m')_{v,v'} $\label{line:send-polluted-start}}
    		{%Pre
                \State $\pr'\in N_{mesh}(\pr)$
                \State $\act{isPolluted}[m] = true$
    		}
    		{%Eff
                \State $m'\gets m$
    			%\State $\act{doneSent}[m] \gets \act{doneSent}[m]\cup\{v'\}$
    		}\EndOutput

            \Statex
            
            \State\Comment{received a notification of a polluted peer}
            \Input{$\act{receive-polluted}(m)_{v',v}$\label{line:rcv-polluted-start}}
            {%Eff
                \State$\act{quaPeers}[m]\gets \act{quaPeers}[m]\cup\{\pr'\}$
                \State$\act{malPeers}[m]\gets \act{malPeers}[m]\setminus\{\pr'\}$
            }\EndInput

        }\EndPart%PART TRANSITIONS
    }
    \end{multicols}
\end{algorithmic}
\caption{\optpp{} Rugby Protocol: Signature and Transitions.
%Protocol for gossiping at any node $v, v \in V$ in the graph $G = (V, E)$.
}
\label{algo:optp2p:rugby}
\end{algorithm*}

\subsection{IOA Specification}
As in Section~\ref{sec:optp2p:gossip}, the rugby protocol is also specified using the 
IOA notation as presented in Algorithm~\ref{algo:optp2p:rugby}. Here, we describe the state variables and the transitions of the protocol.

\paragraph{\bf{State Variables.}} Every node $v$ maintains the following state variables:

%\begin{enumerate}
    %\item[] 
    % \myparagraph{$\act{msgHash}$:} a key-value map where the keys are message ids and the 
    % value is the hash of the message $id$ as computed by the publisher.
    
    %\item[]  
    \myparagraph{$\act{malCodes}[m]$:} 
    %a key-value map where the keys are message $id$s and the values are pairs of pear ids and coded shards for the same message. 
    This variable 
    maintains the polluted shards discovered for each message $m$ by node $\pr$.
    
    %\item[] 
    \myparagraph{$\act{malPeers}[m]$:} 
    %a key-value map where the keys are message $id$s and the values are sets of peer identifiers. 
    Sets of peer identifiers 
    %where each set contains the identifiers we 
    that either 
    detected that they propagated a polluted shard for $m$, or we received an alert 
    %this information 
    from a neighbor. This set contains only nodes for which we have not received a self-pollution message. 
    
    %\item[] 
    \myparagraph{$\act{quaPeers}[m]$:} 
    %a key-value map where the keys are message $id$s and the values are sets of peer identifiers. 
    Sets of peer identifiers which include the peers that informed node
    $\pr$ that they are polluted. 
    % again key-value maps where the keys are nodes $v' \in N(v)$ and the values are a set of vectors or some representation of the basis of a subspace.
    
    %\item[] 
    \myparagraph{$\act{quaShards}[m]$:} set of shards that are received from quarantine peers
    in $\act{quaPeers}$ for message $m$. The integrity of those shards need to be checked before 
    added in the $\act{shardSet}[m]$ along with non-polluted shards. 
    
    %\item[] 
    \myparagraph{$\act{isPolluted}[m]$:}  
    %a key-value map where the keys are message $id$s and the value is 
    A boolean indicating whether node $\pr$ detected that the message $m$ 
    is polluted and cannot be decoded correctly.
%\end{enumerate}

\paragraph{\bf{Transitions.}}
To accomodate the pollution detection and mitigation described in this section 
we had to apply modifications in the transitions of the \optpp{} protocol 
as presented in Algorithm~\ref{algo:optp2p} (as they appear in {\color{olive} olive}).
%More precisely, 
Here are the changes in the transitions of Algorithm~\ref{algo:optp2p}.

% \myparagraph{{\bf \act{generate-encode-data}}:} before completing this action,
% the publisher $\pr$ generates and signs the hash of message $id$. This hash is 
%stored in $\act{msgHash}[id]$.

\myparagraph{{\bf \act{send-shard}}:} a node $\pr$ can now trigger the 
\act{send-shard} action for a message $m$ only if $\pr$ did not discover 
that $m$ is polluted. Moreover, along with the shard data, $\pr$ needs to sign 
the shard to be send as the generator of that shard.

\myparagraph{{\bf \act{receive-shard}}:} this action went through the most changes. 
%This is because w
When a node $\pr$ receives a shard, it checks whether the 
signature included can be verified. 
%in the received message is invalid (using the hash signature), or
Then it examines 
if the received shard 
was sent by a peer $\pr'$ already known to be malicious or quarantined. If 
$\pr'$ does not belong to any of those sets, then 
%none of the above holds then $\pr$ simply adds 
the received shard is added in the $\act{shardSet}$ along with the identifier of the sender 
and the attached signature. If the $m$ is not polluted, it may then generate a new shard 
and add it in the $\act{sendBuffer}$ to be sent to its neighbors.
% and sets $\act{msgHash}[id]$ to be equal to the received hash. Note that this hash is signed by the publisher node, so that no malicious peer can decode and sign a polluted value. 
In case
$\pr'$ is in the $\act{quaPeers}[id]$ set, i.e., $\pr$ received a message that 
$m$ is polluted in $\pr'$, we add the shard into the $\act{quaShards}[m]$ to verify whether 
the received shard is polluted or not. 
If on the other hand $\pr'$ is not in $\act{quaPeers}[m]$ but it is known 
to be a malicious node, i.e., $\pr'$ is in $\act{malPeers}[m]$, then we just discard 
the received shard.

\myparagraph{{\bf \act{decode-msg}}:} when we decode a message we just check if the 
hash of the decoded value matches the hash we maintain for the message $m$. If
a node $\pr$ can decode a message $m$ then it marks the message as non-polluted. 

Given the changes in Algorithm~\ref{algo:optp2p} we can now describe the new transitions
in Algorithm~\ref{algo:optp2p:rugby}.

\myparagraph{{\bf \act{pollution-discovery}}:} this action is triggered when node $\pr$
can decode correctly but yet it discovers a set of $k$ shards in its $\act{shardSet}[m]$ 
of which the hash of the decoded outcome 
is different than the hash of the message $m$. Node $\pr$ then 
removes one random shard from the set of shards that were used to decode
correctly and repeats in a brute force fashion to discover which 
shards in $S$ may be polluted. Every polluted shard is
added in $\act{malShards}[m]$ set and the peer that generated the shard 
in the $\act{malPeers}[m]$ set. 

\myparagraph{{\bf \act{send-alert}}:} the action is triggered when there 
are polluted codes in the $\act{malShards}[m]$ set for a message $m$. 
The node $\pr$ sends the polluted shard along with 
the id of the source and the signature of the shard to the full-neighbors.

\myparagraph{{\bf \act{receive-alert}}:} upon receiving an alert, node $\pr$
checks if the shard signature is valid and then (if possible) whether the shard is polluted. 
If it is verified it adds 
the sender of the shard $\pr_b$ in the malicious set $\act{malPeers}[id]$. 
Notice that $\pr$ does 
not need to maintain the polluted shard as it will not propagate it further 
to avoid network congestion. Moreover, if $\pr_b$ is placed in the $\act{malPeers}[m]$
if it is not already in the $\act{quaPeers}[m]$.
%, and hence informed $\pr$ that it is polluted.
%, then it does not move the peer in the .

\myparagraph{{\bf \act{self-isolate}}:} a node that has more than 
$k$ elements in its $\act{shardSet}[m]$ yet it cannot decode $m$ correctly, 
it proclaims its-self as polluted. As seen above the node stops sending shards for 
message $m$ while in a polluted state. The polluted state is removed 
when $\pr$ manages to decode $m$ correctly. 

\myparagraph{{\bf \act{send-polluted}}:} a node $\pr$ sends this 
message when polluted to the full-neighbors to inform them that 
it is going to stop sending shards for $m$ until it successfully decodes it. 
This will  
allow them to check any shard they receive for $m$ from $\pr$.

\myparagraph{{\bf \act{receive-polluted}}:} a node receiving the polluted 
message from $\pr'$, adds $\pr'$ in their quarantine peers set $\act{quaPeers}$ 

\myparagraph{{\bf \act{check-shards}}:} as seen in the changes of Algorithm~\ref{algo:optp2p}, 
if a node $\pr$ receives a shard from 
a quarantine peer on a message $m$, it adds the shard 
in the $\act{checkShards}[m]$ set. This action is trigerred if there is 
a shard to be checked. If the shard can be used to decode correctly then 
the sender $\pr'$ is removed from the quarantine and the shard is stored
in $\act{shardSet}[m]$.

\section{Experimental Evaluation}
\label{sec:optp2p:evaluation}

\nn{In this section we present the experimental evaluation of \optpp{}, and its performance comparison to that of \gossipsub{}. 
Our experiments include both
%which includes both 
simulation results and real-world deployments. }
% results, as well as real world experiments. We compare 
% the performance of \optpp{} with that of \gossipsub{}.

\subsection{\optpp{} Simulation Results}

\begin{figure*}[ht!]
    \includegraphics[width=\textwidth]{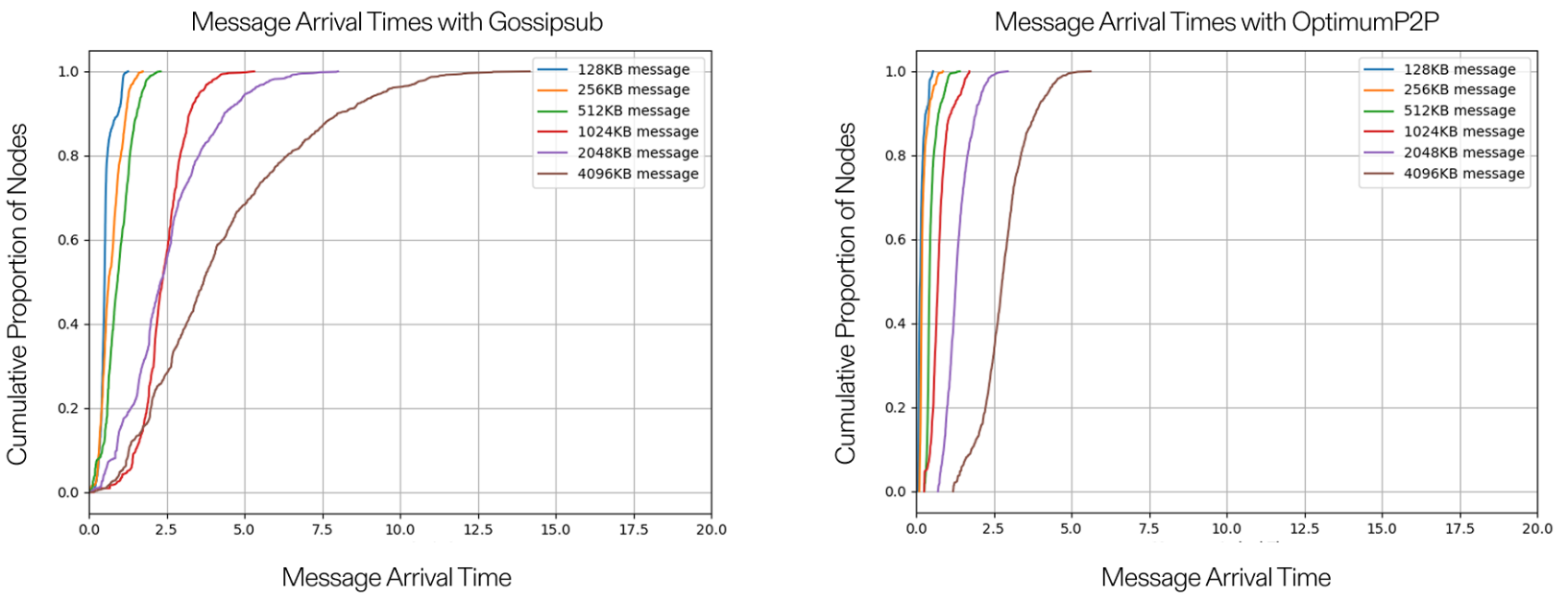}
  \caption{Comparison of latency between  OptimumP2P and Gossipsub by message size.}
  \label{fig:simulation}
\end{figure*}

%TODO:: copy paste need a correct writting and citation
We used the Ethereum tool Ethshadow\footnote{https://ethereum.github.io/ethshadow/} to simulate gossip in an Ethereum-like network. 
%This tool is built on the general-purpose simulation library Shadow \footnote{https://shadow.github.io/}.
We built off the work of prior Ethereum research \cite{ethresearch-rlnc-artice}, running the same simulations as them with 1,000 nodes, 20\% of which have incoming/outgoing bandwidth of 1Gbps/1Gbps and 80\% of which have 50Mbps/50Mbps. The publisher always has 1Gbps/1Gbps in order to get consistent simulation results. The latencies between pairs of nodes are based on real-world geographic locations.
We first ran a simple experiment, in which a single publisher publishes a single message. We varied message sizes from 128KB to 4096 KB, and observed significantly faster arrival times for all message sizes, as shown below. We remark that our observed performance of Gossipsub matches the results in \cite{ethresearch-rlnc-artice}, 
which give us confidence in our reproducibility. 
% We then ran simulations in which a single publisher published multiple messages (up to 64), with each message having a size of 128 KB. Once again, we observed notably faster arrival times in all cases.

\begin{figure*}[ht!]
    \includegraphics[width=\textwidth]{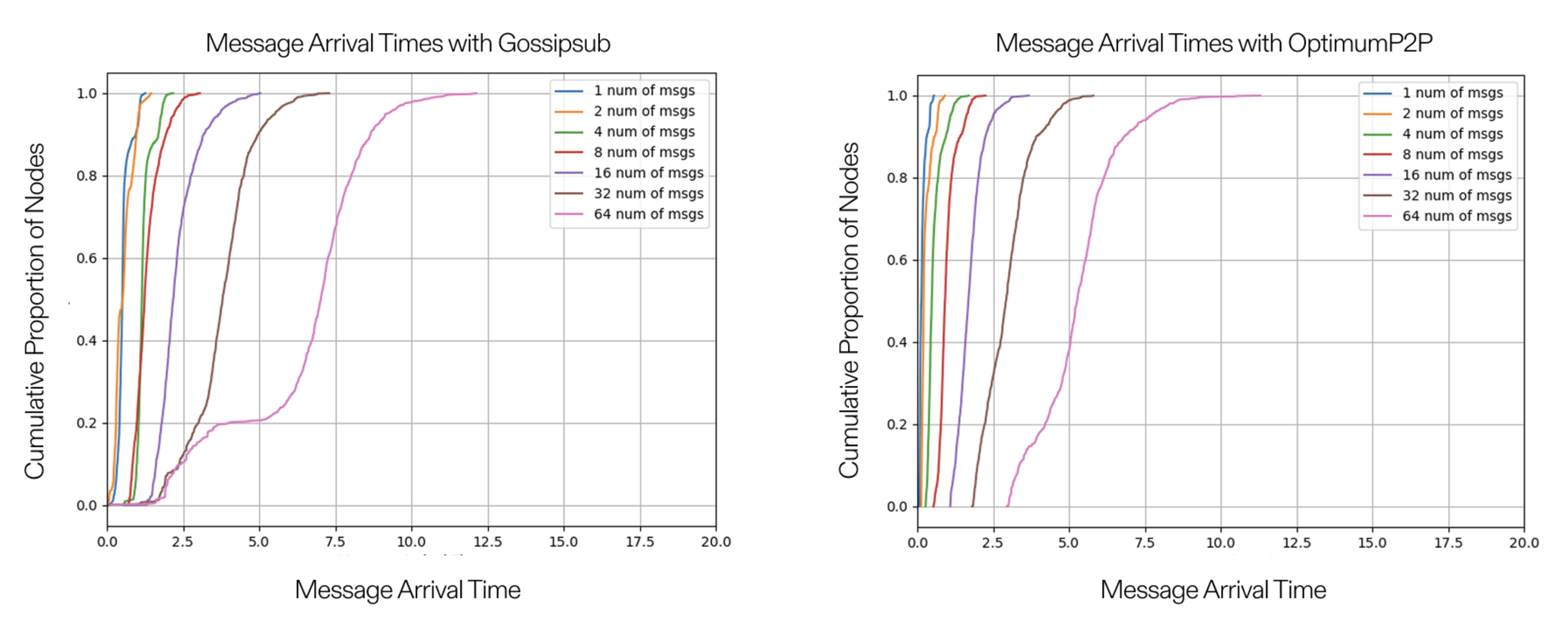}
  \caption{Comparison of latency between  OptimumP2P and Gossipsub by publish rate.}
  \label{fig:simulationrate}
\end{figure*}

\nn{We also ran simulations in which a single publisher published multiple messages (up to 64), with each message having a size of 128 KB. The results appear in Figure \ref{fig:simulationrate}. Once again, we observed notably faster arrival times in all cases.}

\subsection{\optpp{} Real-World Experiments}

\begin{figure}[h!]
\begin{center}
        \includegraphics[width=0.5\textwidth]{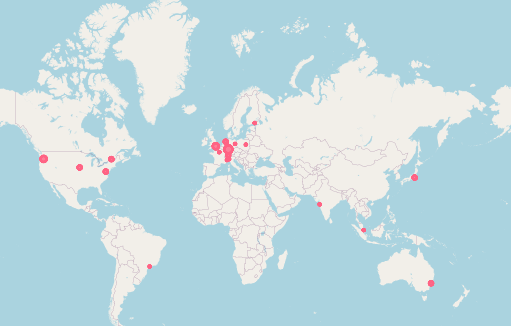}
  \caption{Geographic distribution of the 36 nodes used in each protocol.}
  \label{fig:distribution}
\end{center}
\end{figure}

We performed side-by-side A/B testing of \gossipsub{} and \optpp{}, each deployed across 36 geographically distributed identical nodes in Google Cloud Platform (GCP) data centers (Figure \ref{fig:distribution}), roughly mirroring the distribution of Ethereum validator nodes. 
% the two protocols under identical conditions, using 36 nodes for each. We conducted tests on two networks: one running \gossipsub{} and the other running \optpp{}, each deployed across 36 geographically distributed nodes in Google Cloud Platform (GCP) data centers (Figure \ref{fig:distribution}), roughly mirroring the distribution of Ethereum validator nodes. 
In each test, nodes propagated large data blobs, simulating transaction blocks. A randomly selected node initiated each gossip round, and propagation was deemed successful once at least 95\% of nodes received the message. 
% We measured delivery \textit{latency}, i.e. the time between publish and gossip completion, 
% and reliability, i.e. number of published messages that completed, among nodes in similar regions
% The time inverval between the publish event and the event that completes the gossip is called \emph{latency of gossiping} for that message. 
We varied two main parameters: (i) the message size (from 4MB up to 10MB blobs), and (ii) the publish rate (ranging from isolated single-block sends to rapid bursts up to several messages per second). Both protocols were pushed to carry \textbf{100 messages} per run in some high-load scenarios to observe behavior under message bursts.

The performance metrics recorded include the \textit{propagation latency}, i.e, the time for 95\% 
%all 
of the nodes to receive and reconstruct the message, and the \textit{delivery ratio}, i.e. fraction of nodes that obtained the message within a fixed time-bound. We also tracked the average per-message delay and its variance to gauge stability. 

\begin{figure}[h!]
    \includegraphics[width=0.5\textwidth]{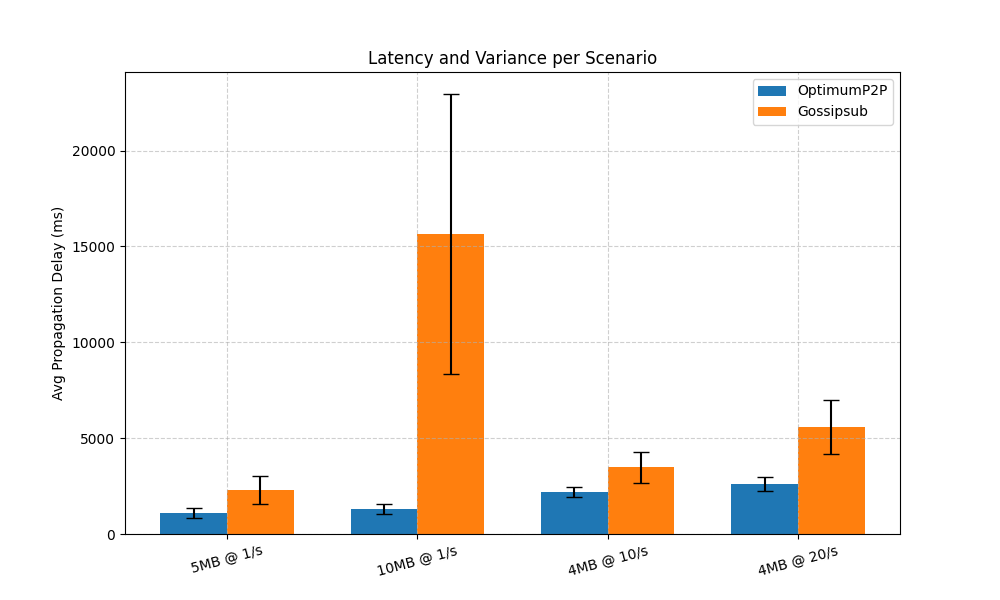}
  \caption{Latency and Variance at Varing Publishing Rates}
  \label{fig:propagation}
\end{figure}

\myparagraph{\textbf{Propagation Latency and Scalability}:}
Figure \ref{fig:propagation} presents the average end-to-end propagation delay for 10MB messages under increasing publish rates, comparing RLNC-based \optpp{} and \gossipsub{}. Lower bars indicate faster delivery.

At \textbf{1 msg/s}, both protocols achieve sub-second latencies; however, \gossipsub{} exhibits a slightly higher delay (~1.0s) than \optpp{} (~0.8s) due to the inefficiencies inherent in gossip-based redundancy and bandwidth usage.

At \textbf{10 msg/s}, \gossipsub{}’s delay increases to ~2.5s, while \optpp{} remains consistently lower at ~1.5s. Under the highest rate of 20 msg/s, \gossipsub{}’s latency sharply degrades to ~4.0s, signaling network saturation and queuing delays. In contrast, \optpp{} maintains delivery within 1.8–2.0s.

These results demonstrate that \textbf{RLNC enables superior scalability} in terms of propagation latency. The coded approach makes more efficient use of network capacity, avoiding redundant transmissions, whereas \gossipsub{}'s performance significantly declines in situations with bursty, high-throughput demands.

We further evaluated scalability under increasing message sizes at a fixed publish rate of \textbf{1 msg/s} testing both 5MB and 10MB payloads.
%
%\begin{itemize}
    %\item 
    For 5MB blocks, \optpp{} achieved 100\% delivery with an average latency of 1116ms (std = 262ms). Gossipsub delivered 99/100, with a delay of 2293ms and higher variance (std = 712ms).
    %\item 
    For 10MB blocks, \gossipsub{}’s performance deteriorated sharply—only 84/100 messages were delivered, with an average delay of 15.6s and significant variability (std = 7.3s). \optpp{} again delivered 100\%, with latency held to 1302ms and low deviation (std = 270ms).
%\end{itemize}

These results demonstrate that \optpp{} remains strong and efficient as message sizes grow, consistently achieving latencies. In contrast, Gossipsub becomes unreliable under larger payloads, succumbing to congestion and protocol overhead.

\begin{figure}[h!]
    \includegraphics[width=0.5\textwidth]{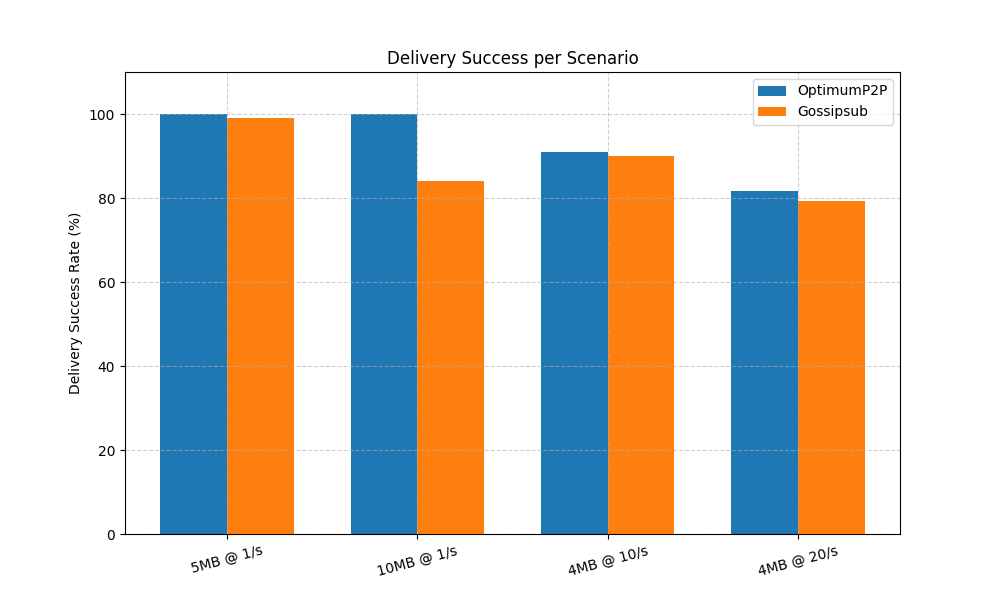}
  \caption{Delivery Success Rate vs Publishing Rates and Message Sizes}
  \label{fig:delivery}
\end{figure}

\myparagraph{\textbf{Throughput and Delivery Success Rate}:}
Figure \ref{fig:delivery} presents the delivery success rate defined as the percentage of messages delivered network wide under increasing publish rates. The results show a consistent advantage for \textbf{RLNC-based OptimumP2P} over \textbf{Gossipsub}, especially under high-throughput conditions.

At \textbf{1 msg/s}, \gossipsub{} and \optpp{} both deliver nearly 100\% of messages, and 
no significant reliability issues are observed at this baseline rate.
At \textbf{10 msg/s}, \gossipsub{} drops to \textbf{approx. 95\%} delivery, 
likely due to packet loss or transient overload in the gossip mesh,
while \optpp{} maintains a \textbf{100\%} delivery rate, 
benefiting from RLNC’s redundancy elimination and network coding.
At \textbf{20 msg/s}, \gossipsub{} falls to \textbf{approx. 80\%} delivery, with approximately \textbf{1 in 5 messages lost}, likely due to congestion, buffer overflows, or gossip suppression.
\optpp{} continues to deliver \textbf{~99–100\%} of messages across the network, with minimal loss despite the high throughput.

% \begin{itemize}
%     \item At \textbf{1 msg/s}:
%         \begin{itemize}
%             \item Gossipsub and OptimumP2P both deliver nearly 100\% of messages.
%             \item No significant reliability issues are observed at this baseline rate.
%         \end{itemize}
%     \item At \textbf{10 msg/s}:
%     \begin{itemize}
%         \item Gossipsub drops to \textbf{approx. 95\%} delivery, likely due to packet loss or transient overload in the gossip mesh.
%         \item OptimumP2P maintains a  \textbf{100\%} delivery rate, benefiting from RLNC’s redundancy elimination and network coding.
%     \end{itemize}
%     \item At \textbf{20 msg/s}:
%     \begin{itemize}
%         \item Gossipsub falls to \textbf{approx. 80\%} delivery, with approximately \textbf{1 in 5 messages lost}, likely due to congestion, buffer overflows, or gossip suppression.
%         \item OptimumP2P continues to deliver \textbf{~99–100\%} of messages across the network, with minimal loss despite the high throughput.
%     \end{itemize}
% \end{itemize}

These findings underscore the throughput advantage of RLNC-based dissemination. \optpp{}’s ability to deliver coded fragments and recover full messages from partial data ensures robust delivery even under stress. In contrast, \gossipsub{}’s reliance on full-message relays makes it vulnerable to packet drops and network saturation during bursts.

\begin{figure}[h!]
    \includegraphics[width=0.5\textwidth]{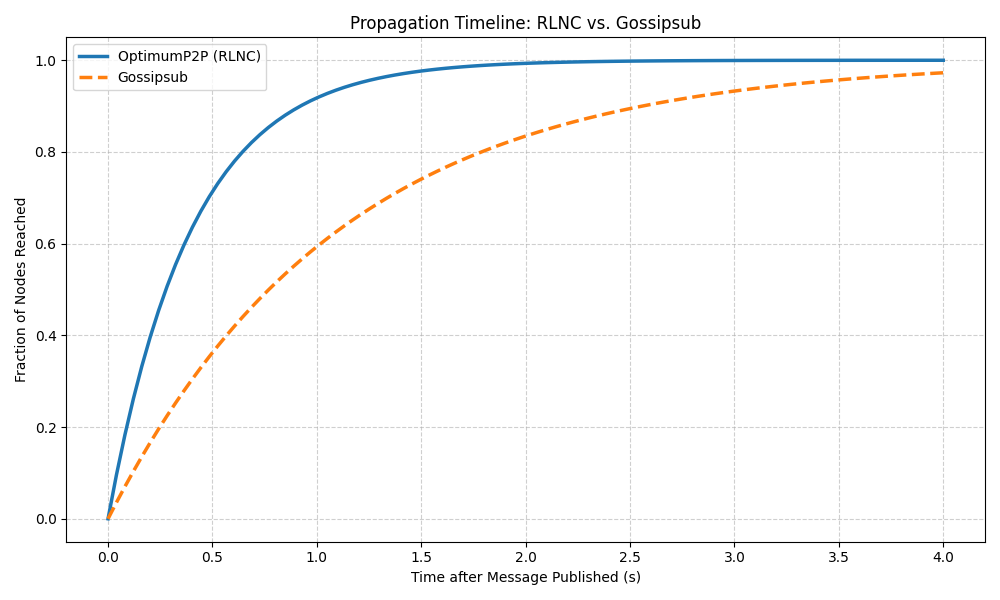}
  \caption{Propagation Stability of the protocols}
  \label{fig:stability}
\end{figure}

\myparagraph{\textbf{Propagation Stability and Delay Variance}:}
Figure \ref{fig:stability} illustrates the \textbf{mean propagation delay} and \textbf{standard deviation} across nodes for different publish rates. These error bars reveal stable or erratic delivery times under varying network loads.

%\begin{itemize}
    %\item 
    Across all rates, \textbf{\optpp{}} exhibits \textbf{consistently low variance}, with standard deviation tightly bounded around \textbf{±0.2–0.3 seconds}. This suggests %Suggests 
    a highly predictable propagation process, where most nodes receive messages within a limited time window.
    
    %\item 
    \gossipsub{}, in contrast, shows \textbf{significantly higher delay variability}, especially at \textbf{higher publish rates}. At \textbf{20 msg/s}, the standard deviation grows to nearly \textbf{±0.9 seconds}, indicating that some nodes receive messages much later than others. This inconsistency stems from \gossipsub{}’s multi-hop gossip structure, which can lead to inconsistent dissemination and redundant retransmissions.
    
    %\item 
    The \textbf{lower variance} in \optpp{} designs is due to its \textbf{pipeline-friendly, parallel dissemination} using RLNC. Coded fragments are spread uniformly and decoded incrementally, reducing reliance on any single route or node.
%\end{itemize}

These results demonstrate that \optpp{} provides faster and more predictable delivery, critical for latency-sensitive applications like block propagation in Blockchain systems like Ethereum. In contrast, \gossipsub{}’s performance becomes erratic under load, undermining its suitability for time-critical scenarios.

\section{Conclusions}
\label{sec:conlcude}
%In this work-in-progress (WIP), 
We presented \optpp{}, a gossip protocol that utilizes Random Linear Network Coding (RLNC) to enhance the speed of information propagation in peer-to-peer (p2p) networks. By leveraging the properties of recoding in RLNC, \optpp{} outperforms current solutions by achieving faster network coverage and reducing message duplication. 
This enables \optpp{} to reach peers in the network faster while preserving network bandwidth. In turn, \optpp{} provides clear benefits to distributed solutions that require information propagation among a set of network nodes, such as block propagation in modern blockchain solutions. The performance
gains are evident from our experimental evaluation where we compare \optpp{} with the state-of-the-art
\gossipsub{} implementation, both in simulation and real-world setups. 
\vspace{0.5cm}

\myparagraph{\textbf{Acknowledgments}:} The computing resources for the deployment of the experimental work in this paper are supported in part by the Startups Cloud Program by Google.
% 

% The next steps involve theoretically analyzing the \optpp{} algorithm, both in terms of correctness and performance. Furthermore, we plan to deploy \optpp{} within libp2p2 and conduct thorough experimentation and comparison with existing solutions. 

%%%
%%% BIBLIOGRAPHY
%%%
\bibliographystyle{acm}
\bibliography{Galois-Gossip/ARXiV/GGBiblio}

%\begin{verbatim}
%https://github.com/libp2p/specs/blob/master/pubsub/gossipsub/gossipsub-v1.0.md   
% added to bib as libp2p-pubsub and libp2p-gossip-sub
% https://docs.libp2p.io/concepts/pubsub/overview/
%\end{verbatim}

\clearpage

%\section{Appendix}
%\appendix

\end{document}